\documentclass[12pt]{article}
\usepackage{amsmath}
\usepackage{natbib}
\usepackage{algorithm}
\usepackage{algpseudocode}
\usepackage{lscape}
\usepackage{setspace}
\usepackage{authblk}
\usepackage{graphics,graphicx}
\usepackage{appendix}
\usepackage{wrapfig}
\usepackage{caption}
\usepackage{subcaption}
\usepackage{url}
\usepackage{amsmath,amssymb,amsthm}
\usepackage{bm}
\usepackage{graphicx}
\usepackage{booktabs}
\usepackage{comment}
\usepackage{enumitem}
\usepackage{multirow}
\usepackage{array}
\usepackage{float}
\usepackage{xcolor}
\usepackage{rotating}
\setlist{nolistsep}

\makeatletter
\renewcommand\@biblabel[1]{#1.\hfill}
\makeatother

\DeclareUnicodeCharacter{0308}{HERE!HERE!}

\newcommand{\blind}{1}

\date{}

\begin{document}

\def\spacingset#1{\renewcommand{\baselinestretch}%
{#1}\small\normalsize} \spacingset{1}


\if1\blind
{
  \title{\bf Design and Analysis Strategies for Pooling in High Throughput Screening: Application to the Search for a New Anti-Microbial}
    \author[1,2,3]{Byran J. Smucker\thanks{Corresponding author (smucker6@msu.edu)}
}
    \author[3]{Benjamin Brennan}
    \author[4,5]{Emily Rego}
    \author[6,7]{Meng Wu}
    \author[6]{Zhihong Lin}
    \author[4,5]{Brian M. M. Ahmer}
    \author[6,7]{Blake R. Peterson}
    
    \affil[1]{Department of Epidemiology \& Biostatistics, College of Human Medicine, Michigan State University, East Lansing, MI, USA}
    \affil[2]{Henry Ford Health + Michigan State University Health Sciences, Detroit, MI, USA}
    \affil[3]{Department of Public Health Sciences, Henry Ford Health, Detroit, MI, USA}
    \affil[4]{Department of Microbiology, The Ohio State University, Columbus, OH, USA}
     \affil[5]{Department of Microbial Infection and Immunity, The Ohio State University, Columbus, OH, USA}
    \affil[6]{The Ohio State University Comprehensive Cancer Center – Arthur G. James Cancer Hospital and Richard J. Solove Research Institute, Columbus, OH, USA}
    \affil[7]{Division of Medicinal Chemistry and Pharmacognosy, The Ohio State University, Columbus, OH, USA}

  \maketitle
} \fi

\if0\blind
{
  \bigskip
  \bigskip
  \bigskip
  \begin{center}
    {\LARGE\bf Design and Analysis Strategies for Pooling in High Throughput Screening: Applications to the Search for a New Anti-Microbial}
\end{center}
  \medskip
} \fi

\bigskip
\begin{abstract}
A major public health issue is the growing resistance of bacteria to antibiotics. An important part of the needed response is the discovery and development of new antimicrobial strategies. These require the screening of potential new drugs, typically accomplished using high-throughput screening (HTS). Traditionally, HTS is performed by examining one compound per well, but a more efficient strategy pools multiple compounds per well. In this work, we study several recently proposed pooling construction methods, as well as a variety of pooled high-throughput screening analysis methods, in order to provide guidance to practitioners on which methods to use. This is done in the context of an application of the methods to the search for new drugs to combat bacterial infection. We discuss both an extensive pilot study as well as a small screening campaign, and highlight both the successes and challenges of the pooling approach.
\end{abstract}

\noindent%

\spacingset{1.9} 

\section{Introduction}\label{sec:introduction}

Pathogens are becoming increasingly resistant to antibiotics, and this fact has alarming implications for public health including increased mortality and associated healthcare costs \citep{tacconelli2017global, manesh2021rising}. The United Nations estimates that deaths due to antimicrobial resistance will be 10 million per year by 2050 \citep{coque2023bracing}. 
For a specific example, the typhoidal strains of \textit{Salmonella enterica} cause typhoid fever resulting in more than 100,000 deaths per year \citep{mogasale2014burden, antillon2017burden}, and non-typhoidal strains are a leading cause of death from food-borne illness in the United States \citep{scallan2011foodborne} and a contributor to the mortality in developing nations due to diarrhea \citep{kotloff2013burden, pires2015aetiology}. Because of the surge in antibiotic-resistant typhoidal and non-typhoidal \textit{Salmonella enterica} strains, the CDC and the WHO have called for the development of new drugs to address this need \citep{tacconelli2017global}. 

In response to this problem, a new anti-microbial strategy has been suggested \citep{schwieters2025mtld} with potential in many bacteria including \textit{Salmonella enterica}. An important part of developing that strategy is drug discovery, and the initial phase of this search is accomplished via high-throughput screening (HTS). High-throughput screening is used widely in drug discovery, chemical biology, and many other scientific and industrial domains. HTS involves the placement of specified compounds into wells in 384- or even 1536-well plates. The plates are then assayed and checked for a desired indication. Normally, a single compound is placed in each well, but various authors have suggested and shown that pooling multiple compounds in each well can improve throughput and/or statistical efficiency. Pooling has been controversial and contested, with both successes \citep[e.g.,][]{wilson1996deconvolution,  snider1998screening, motlekar2008evaluation, elkin2015just} and cautions \citep{chung1998screen, feng2006synergy} in the literature. Recently, however, there have been reports of several successful pooled screening procedures \citep{ji2023target, liu2024scalable, smucker2025large}. Older methodologies, using approaches like orthogonal pooling \citep{kainkaryam2009pooling} or poolHiTS \citep{kainkaryam2008poolhits}, neither constructed their pools nor analyzed them using statistical methods, while the new approaches use statistical design ideas for pool construction \citep[][]{ji2023target, smucker2025large} and statistical regularization for analysis \citep[][]{ji2023target, liu2024scalable, smucker2025large}.  

In this work, we add to this emerging pooling literature in several ways. First, we describe in some detail a particular application related to a search for antimicrobials discussed at the outset (Section \ref{sec:application_description}). In Section \ref{sec:methods}, we describe a number of existing pool construction methods and pooled HTS analysis methods, and make a set of extensive comparisons between them in Section \ref{sec:comparisons}. Most of the methods we consider are from the literature, but we propose a new Lasso thresholding method that exploits our knowledge of effect directions. We also discuss a secondary analysis method to address a problem in these types of screens: they often produce too many false positives, which  consume a large amount of resources. The secondary criterion severely reduces the number of compounds to be validated, while still detecting large effects. In Section \ref{sec:application} we then present an extensive description of the pilot study (Section \ref{sec:pilot_results}) used to establish pooling as a viable approach in this setting, as well as results from an initial screen (Section \ref{sec:screening_results}) which identified several promising compounds while reducing the number of considered false positives. We finish with a Discussion in Section \ref{sec:discussion}.

\section{The Screening Problem} \label{sec:application_description}

Schwieters et al. \citep{schwieters2025mtld} report that the enzyme mannitol-1-phosphate 5-dehydrogenase (MtlD) offers antimicrobial potential in many bacteria. This enzyme converts the compound mannitol-1-phosphate to fructose-6-phosphate. When \textit{mtlD} mutant bacteria are provided mannitol, mannitol-1-phosphate accumulates, intoxicating the bacterium leading to reduced growth and attenuated virulence in animal models. Thus, the goal is to identify a compound that inhibits MtlD in the presence of mannitol. Using a wild-type bacterium, inhibition of MtlD in the presence of mannitol will result in lack of growth. However, numerous compounds will inhibit the growth of a wild-type bacterium for reasons unrelated to MtlD inhibition. To eliminate these from consideration, a parallel screen is used in which the bacterium cannot be harmed by a MtlD inhibitor (because it is a \textit{mtlA} mutant and cannot form mannitol-1-phosphate). Thus, the screening problem is to find a drug which inhibits growth of the wild-type (WT) bacteria but not of the \textit{mtlA} mutant (MUT) when assayed against both. We call such a drug a \emph{true hit}. A drug which inhibits both WT and MUT is called a \emph{pseudo-hit}. Our goal is to identify true hits.

Previous single-replicate, one-compound-one-well (OCOW) screening of 10{,}000 compounds for this system yielded 140 that appeared to inhibit WT, and 40 of those did not appear to inhibit MUT. However, upon retest in duplicate, none of the 140 were validated as true hits. Contemplating a similar screen scaled up to hundreds of thousands of compounds, the investigators realized that even a false positive rate of 1\% would be extremely expensive. This led them to consider pooling as a more efficient alternative, since it is a way to observe each compound more than once while using more compounds than wells. This yields the potential of larger true positive rates with the same or smaller false positive rates. The researchers agreed that eight compounds per pool was feasible. From their previous testing, they expect the set of true hits to be extremely sparse within the space of compounds they anticipate searching. Statistically, this does have an advantage: Interactions between compounds are not expected, unless two or more true hits end up in the same pool. Because of the level of sparsity, this is unlikely.

Despite its promise, pooling also presents challenges. Logistically, the compounds, stored in source plates, must be transferred to target plates in pools dictated by the experimental design. This is accomplished via automated liquid handling machines, but constructing the pooled plates is still more time-consuming than OCOW. It also requires careful preparation and communication between the statistician and the personnel implementing the design. We revisit this screening problem in Section \ref{sec:application}.

\section{Design and Analysis Methodologies} \label{sec:methods}

In addition to demonstrating an application of pooling in HTS and the associated data and statistical challenges, two additional objectives of this work are to (1) compare pooled HTS design strategies; and (2) compare pooled HTS analysis strategies. A basic question we seek to learn about: If one wishes to undertake a pooled high-throughput screen, what design and analysis method will be most effective?

In this section, we describe the designs and analysis methods we consider. Throughout, we use the designs and analysis methods to estimate the main effects model, which we assume to be true: 
\begin{align}
\mathbf{y}=\beta_{0}\mathbf{1} + X\boldsymbol\beta + \boldsymbol\epsilon, \label{eq:model} 
\end{align}
where $\mathbf{y}$ is $n \times 1$, $X \in \{-1,1\}^{n \times k}$, $\boldsymbol\beta=(\beta_{1},\ldots,\beta_{k})^{T}$ and $\boldsymbol\epsilon \sim N(\mathbf{0},\sigma^{2}I)$ with $\mathbf{0}$ an $n$-vector and $I$ the $n\times n$ identity matrix. 

\subsection{Designs} \label{sec:design_methods}

Following the model above, we assume $n$ wells to study $k$ compounds, with $n<k$. The design $X$ is coded such that entry $ij$ is $-1$ if the $j^{th}$ compound is absent in the $i^{th}$ well, and $+1$ if the compound is present. We represent the size of the $i^{\text{th}}$ pool by $c_{i}$, with $c_{\text{min}}$ and $c_{\text{max}}$ denoting the smallest and largest pool size in a design, respectively. If $c_{\text{min}}=c_{\text{max}}$, let $c$ denote the common size across all pools. Similarly, the number of times the $j^{\text{th}}$ compound appears in the design is $a_{j}$, with $a_{\text{min}}$, $a_{\text{max}}$, and $a$ defined analogously. In our study, we primarily consider three types of designs: (1) the Constrained Row Screening designs of Smucker et al. \citep{smucker2025large}, based on a criterion from the supersaturated design literature; (2) the matrix-augmented pooling strategy of Ji et al. \citep{ji2023target}, using ideas from the compressed sensing literature; and (3) a semi-random construction approach from Liu et al. \citep{liu2024scalable}. 

\subsubsection{Constrained Row Screening Designs} \label{sec:CRowS}

The Constrained Row Screening (CRowS) designs of Smucker et al. \citep{smucker2025large} are constructed using the $UE(s^2)$-criterion from the supersaturated design literature. In this design formulation, the maximum pool size $c_{max}$ is specified and this translates to a constraint on the number of $+1$'s in each row. Given the $\{-1,1\}$ coding, this translates to CRowS designs respecting the constraint $\sum_{j=1}^{k} x_{ij} \leq 2c_{max} - k$ for all rows, $i=1,2,\ldots,n$. The $UE(s^2)$ criterion is $UE(s^2)=\sum_{i<j}s_{ij}^{2}/k(k+1)$, where $s_{ij}$ is the $ij^{th}$ element of $S=L^{'}L$ and $L=[\mathbf{1},X]$. This criterion attempts to make the off-diagonals of $S$ to be as small as possible, where $S$ captures the amount of similarity between the columns of $L$. Since $c_{max}$ is often much less than $k$, these off-diagonals are typically much larger than 0. Details of design construction and Matlab code can be found in Smucker et al. \citep{smucker2025large} and its supplementary material. 

We have observed that when $c_{max}$ is small relative to $k$, all rows tend to have exactly $c=c_{max}$ $+1$'s; that is, all pools are of size $c$. Proving this observation is still an open question. In our observation, the designs also have that each compound appears $a=a_{max}$ times, if $n$, $k$, and $c$ parameters allow, though this too has not been shown mathematically. Using the methods of \citep{smucker2025large}, CRowS designs such as those used in this paper can be constructed in a few minutes.

\subsubsection{Matrix-Augmented Pooling Strategy} \label{sec:MAPS}

The matrix-augmented pooling strategy (MAPS) \citep{ji2023target} uses ideas from the compressed sensing literature to generate a matrix with relatively small correlations between columns. Instead of constraining the pool size, MAPS requires $a_{\text{min}}$ to be specified; that is, each column must have at least $a_{min}$ $+1$'s.  

Let $U \in \{0,1\}^{n \times k}$ be a design matrix such that an entry of $0$ for entry $ij$ represents compound $j$ absent in pool $i$, while an entry of $1$ represents the compound's presence. MAPS attempts to minimize \[
\begin{aligned}
M 
&= \| U^{\top} U - AI_k \|_F^2
   + \sum_{i=1}^{k} \left( \sum_{j=1}^{n} U_{ij} - \overline{s} \right)^2, \\[6pt]
 \end{aligned}
\]
where $\overline{s} = \frac{1}{k} \sum_{i=1}^{k} \sum_{j=1}^{n} U_{ij}$ and $A$ is a $k \times k$ diagonal matrix with entries $(i, i)$ indicating the number of drugs in the $i$-th pool for $i = 1, \dots, k$. Even using a slightly altered, computationally improved version of this Genetic Algorithm-based procedure, construction of designs such as those used in this paper can take multiple hours on a single machine. Note that $X=2U-\mathbf{1}_{n}\mathbf{1}_{k}$ transforms $U$ into the $\{-1,1\}$ coding, which can be used in the analysis of data generated from MAPS designs. Details and Python code can be found in \citep{ji2023target} and its supplementary material.

\subsubsection{Random Pools} \label{sec:random}

Another approach to design construction is to randomly construct the pools, while respecting a strong $nc=ka$ constraint. This restricts the design sizes and pool sizes available to this method and forces the pool size constraint to be respected. This is the method of \citep{liu2024scalable} to the best of our understanding and because of the strong constraint the designs are quite similar, if not identical to, CRowS designs for combinations of $n$, $c$, $k$, and $a$ that satisfy the constraint. Relaxing the $nc=ka$ constraint can produce trivially bad designs with, for instance, columns consisting only of $-1$'s. Designs created via this method can be constructed quickly, usually in a matter of seconds. 

\subsubsection{Comparing the Methods}

In this work we focus on the three recently proposed pooling methods described in Sections \ref{sec:CRowS}-\ref{sec:random}. CRowS and MAPS are statistically derived and the Random pooling procedure is a reasonable benchmark that is amenable to statistical analysis. Other, non-statistical pooling construction methods, such as poolHiTS \citep{kainkaryam2008poolhits} and Orthogonal Pooling \citep[e.g.,][]{kainkaryam2009pooling}, could also be compared. However, \citep{smucker2025large} already demonstrated that CRowS was clearly superior to both OCOW and poolHiTS. Regarding Orthogonal Pooling, it not only suffers from strict limitations of design sizes, but we also show in the Supplementary Material that it is statistically inferior to CRowS, at least for the simulated settings we explored.

To match the real problem we are working on, we will assume 384-well format, and that for each plate 320 wells are available for pools. In an HTS campaign, hundreds of thousands of compounds may be screened, but the compounds must be separated into groups such that each pooled plate is its own experiment. Typically, a single design is applied to each of many plates, and each plate is analyzed separately. The details of the application are described in Section \ref{sec:application}, but the design sizes we considered and simulated from are provided in Table \ref{tab:design_parms}, along with information about the designs, including maximum and minimum pool size, maximum and minimum number of times each compound appears, and the criterion values $\sqrt{UE(s^2)}$ and $\sqrt{M}$. The design sizes are chosen to reflect the designs we used in the application. 
In Table \ref{tab:design_parms} we notice several things. First, the balanced random designs are quite similar to the CRowS designs. Secondly, the MAPS designs do not respect the row constraint and this can result in $UE(s^2)$-inferior designs (see the $k=960$ case). Finally, as expected the CRowS designs are the best in terms of the CRowS criterion, but in most cases they are also better in terms of the MAPS criterion.

\begin{table}
\centering
\begin{tabular}{|c|c|c|c|c|c|c|}
    \hline 
  $n$ & $k$  & Design Type & $c_{\text{min}}$/$c_{\text{max}}$ & $a_{\text{min}}$/$a_{\text{max}}$ & $\sqrt{UE(s^2)}$ & $\sqrt{M}$ \\
  \hline \hline
  \multirow{2}{*}{\centering 320} &
  \multirow{2}{*}{500}  & CRowS & 8/8 & 2/2 & 312.050 & 134.343 \\
\cline{3-7}
      
 & & MAPS & 3/16 & 2/2 & 312.055 & 145.674 \\
    \cline{3-7}
    \hline \hline
    & &
  CRowS & 8/8 & 4/4 & 304.201 & 133.910 \\
\cline{3-7} 
       \multirow{1}{*}{320} &
  \multirow{1}{*}{640}  &
 MAPS & 1/16 & 4/4 & 304.219 & 148.647 \\
    \cline{3-7}
    & & Random & 8/8 & 4/4 & 304.201 & 135.794 \\
    \hline \hline
  \multirow{2}{*}{320} &
  \multirow{2}{*}{960}  &

  CRowS & 8/8 & 2/3 & 309.426 & 133.866 \\
\cline{3-7}
 & & MAPS & 1/13 & 2/2 & 312.056 & 110.941 \\
\cline{3-7}
\hline \hline
  & & CRowS & 8/8 & 2/2 & 312.050 & 133.866 \\
\cline{3-7}
   \multirow{1}{*}{320} &
  \multirow{1}{*}{1280}  & MAPS & 3/16 & 2/2 & 312.055 & 146.956 \\
\cline{3-7}
  & & Random & 8/8 & 2/2 & 312.050 & 134.0895 \\
\hline
\end{tabular}
\caption{Sizes and parameters for the designs we will compare.}
\label{tab:design_parms}
\end{table}

\subsection{Analysis Methods} \label{sec:analysis_methods}

Recent work \citep{ji2023target, liu2024scalable, smucker2025large} has suggested the use of statistical regularization methods to analyze pooled screens, and in this work we will conduct a detailed comparison of a number of these methods and variations on them, in order to provide guidance to practitioners regarding the statistical analysis of pooled HTS.

We can write a general regularized $\ell_{2}$ estimator, called the elastic net \citep{zou2005regularization}, as:
\begin{align}
    \hat{\boldsymbol\beta} \equiv \text{argmin}_{\boldsymbol\beta}\left(\frac{1}{2n}||\mathbf{y}-X\boldsymbol\beta||^{2}_{2} + \lambda\left[(1-\alpha)\frac{1}{2}||\boldsymbol\beta||^{2}_{2} + \alpha||\boldsymbol\beta||_{1}\right] \right), \label{eq:elastic_net}
\end{align}
where $\lambda$ governs regularization aggression and $\alpha$ the degree to which a Lasso-style penalty \citep[][]{tibshirani1996regression} or a ridge-regression penalty \citep[][]{hoerl1970ridge} is prioritized. In what follows, we follow the supersaturated design literature \citep[e.g.,][]{stallrich2025optimal} and emphasize the Lasso---and in particular a method we describe below called the Gauss-Lasso---but we also use the elastic net as suggested in \citep{liu2024scalable} as well as the nonnegative Lasso as in \citep{ji2023target}. The latter is identical to the Lasso except with the additional constraint that $\boldsymbol\beta \geq 0$.

\subsubsection{Gauss-Lasso} \label{sec:gl}

For the Lasso methods we mention below, we used something we call the Gauss-Lasso, following \citep{smucker2025large} and earlier work. This is a two-stage procedure that obtains Lasso fits, thresholds small estimates, then follows up with another fit using ordinary least squares. Specifically, the Gauss-Lasso consists of the following:
\begin{enumerate}
    \item Center $X$ and $\mathbf{y}$, and scale $X$ so that each column as the same length. Call these quantities $X_{cs}$ and $\mathbf{y}_{c}$.
    \item Obtain Lasso solutions for $\log{\lambda}$ for a range wide enough to fully explore reasonable Lasso estimates. For these simulations, the interval was $-5$ to $\log(\max(|X_{cs}^{T}\mathbf{y}_{c}|)$ in increments of $0.25$. 
    \item Across all $\lambda$, set any parameter estimate to 0 if it is greater than threshold $-\tau$. This means we threshold all positive estimates, since we assume all important effects will be negative. Note that without loss of generality, in the simulations we assume positive effects, in which case we threshold any estimates that are smaller than $\tau$. In the details of the various methods we discuss the choice of $\tau$.
    \item Across all $\lambda$, refit each model using OLS and choose the model with the smallest BIC. The factors in the chosen model are considered hits.
\end{enumerate}

In this work we consider four versions of the Gauss-Lasso. 
To mimic an ideal setting in which we know $\sigma$, we investigate the Gauss-Lasso with $\tau=\sigma/8$ and $\tau=\sigma/4$; in real applications, pilot studies or controls may allow a good estimate of $\sigma$. We also consider the Gauss-Lasso with unknown $\sigma$, following the suggestion of \citep{phoa2009analysis} and \citep{weese2021strategies}, by considering $\tau = r_{1} \times \text{max}(|\hat{\beta}_{\lambda=0}|)$, with
$r_{1}=\{0.1,0.5\}$.

\subsubsection{$\lambda$-specific Gauss-Lasso} \label{sec:lam_specific}

In our setting, hit compounds are inhibitory. Thus, we know that any effect of interest is negative and we can assume that any positive Lasso estimates are noise. Then, for any $\lambda$, we can use this noise as the basis to judge the size of an effect that should be considered a potential hit. More specifically, for a given $\lambda$ we threshold all positive estimates as well as any negative Lasso estimate that is less in absolute value than $\tau_{\lambda}=r_{2} \times \text{max}(\hat{\beta}_{\lambda})$, and we consider $r_{2}=\{0.5,0.7,0.9,1\}$. Note that we use the Gauss-Lasso as described in Section \ref{sec:gl}, except that the third item is performed with the thresholding as described here. Further, the simulations described in the next section actually assume positive effect estimates (instead of negative, as in our application), in which case for a given $\lambda$ we threshold all negative estimates as well as any positive Lasso estimates that are less than $\tau_{\lambda}$.

\subsubsection{Non-Negative Gauss-Lasso}

We add an additional constraint via \textit{glmnet} to ensure all coefficients are greater than or equal to 0, following \citep{ji2023target}. Thus, we use the Gauss-Lasso with this additional constraint, repeating the four Gauss-Lasso screening thresholds listed above in Section \ref{sec:gl}.

\subsubsection{Elastic Net} 

Following \citep{liu2024scalable}, we consider a version of the Elastic Net, where the model is specified in \eqref{eq:elastic_net}. To analyze the pooled HTS data, consider a set of the mixing parameters, $\alpha \in \{0.05, 0.25,0.5,0.75, 0.95\}$. The best model, in terms of $\alpha$ and the regularization parameter $\lambda$, is chosen via 3-fold cross validation. Next, we create 1000 permutations of the outcome and fit the best model to each permutation to create a distribution of estimates for each element of $\boldsymbol\beta$. The estimates for each drug-specific parameter, from the actual data fit to the best model, are compared to the parameter's null distribution. If the actual estimate is far enough in the tail of the null distribution then we deem the drug associated with the parameter a hit. More specifically, for each drug-specific parameter, we compute a p-value like quantity as the average number of times, in our 1000 null permutations, that the drug-specific parameter is larger in absolute magnitude than the corresponding parameter in the best model. These p-values were empirically calculated by $p_k = \frac{1}{1000}\sum_{i=1}^{1000} I(|\hat{\beta}_{k}| <|\hat{\beta}_{i,k_0}| )$, where $\hat{\beta}_{k}$ is the estimate from the best model and $\hat{\beta}_{i,k_{0}}$ is the estimate based on the data from the $i^{\text{th}}$ permutation fit to the best model. This version of the elastic net is similar to that used in \citep{LiWu97}. 

In Table \ref{tab:analysis_methods} we provide a list of the 13 analysis methods we consider. The last column in Table \ref{tab:analysis_methods} will be discussed in Section \ref{sec:analysis_comparisons}.

\begin{table}
\centering
\begin{tabular}{|c|c|c|}
    \hline
  Analysis Method & Tuning Parameter & Average $\log(\text{TPR}/\text{FPR})$ \\
  \hline \hline
  \multirow{4}{*}{\centering Gauss-Lasso} & $\tau=\sigma/8$ & 4.45 \\
  
  & $\tau=\sigma/4$ & 4.56 \\
   & $\tau=0.1 \times \text{max}(|\hat{\beta}_{\lambda=0}|)$ & 4.44 \\
   & $\tau=0.5 \times \text{max}(|\hat{\beta}_{\lambda=0}|)$ & 4.96 \\
  \hline \multirow{4}{*}{\centering $\lambda$-specific Gauss-Lasso} &  $\tau_{\lambda}=0.5 \times \text{max}(\hat{\beta}_{\lambda})$ & 4.84 \\
  & $\tau_{\lambda}=0.7 \times \text{max}(\hat{\beta}_{\lambda})$ & 5.09 \\
  & $\tau_{\lambda}=0.9 \times \text{max}(\hat{\beta}_{\lambda})$ & 5.35 \\
  & $\tau_{\lambda}=1 \times \text{max}(\hat{\beta}_{\lambda})$ & 5.44 \\
  \hline \multirow{4}{*}{\centering Non-Negative Gauss-Lasso} & $\tau=\sigma/8$ & 4.42 \\
  & $\tau=\sigma/4$ & 4.52 \\
  & $\tau=0.1 \times \text{max}(|\hat{\beta}_{\lambda=0}|)$ & 4.42 \\
  & $\tau=0.5 \times \text{max}(|\hat{\beta}_{\lambda=0}|)$ & 4.91 \\ 
  \hline Elastic Net & CV & 3.21 \\
  \hline
\end{tabular}
\caption{Analysis methods to compare, along with the variations due to variations in tuning parameters. The last column provides an overall measure of quality, across the 40 simulation conditions defined by the four effect sizes and 10 design/size combinations in Table \ref{tab:design_parms}, as described in Section \ref{sec:sim}.}
\label{tab:analysis_methods}
\end{table}

\section{Comparing Pool Construction and Analysis Strategies} \label{sec:comparisons}

In Section \ref{sec:design_methods}, we described three different pool construction strategies that have recently been paired with regularization for the design of high-throughput screens, along with several analysis strategies in Section \ref{sec:analysis_methods}. In this section, we describe our simulation approach, and compare first the pool construction methods, and then the screening analysis strategies. Our goal is to provide guidance to practitioners regarding the use of design and analysis strategies in pooled HTS applications.

\subsection{Simulation Framework} \label{sec:sim}

Table \ref{tab:design_parms} shows the design sizes and design types that we studied. We chose the design sizes to be analogous to the designs we used in the pilot study we describe in Section \ref{sec:pilot_results}. We generated data from these designs using model $\eqref{eq:model}$ assuming, without loss of generality, that $\sigma=1$ and $\beta_{0}=0$. We assumed a sparsity level of $\lceil 0.01k \rceil$, which guarantees at least one active factor in each simulation while respecting the expected sparsity of the application at hand. We varied the size of the active effects to be $\beta=\{1,2,3,4\}$, which represents the difference in the mean response when a hit compound is in a well vs. when it is not. The value $\beta$ should not be confused with $\beta_{k}$ which is the parameter value for the $k^{\text{th}}$ compound and if it is a hit its value would be $\beta_{k}=\beta/2$. To evaluate performance, we examined true positive rate (TPR; the proportion of truly active factors detected), false positive rate (FPR; the proportion of truly inactive factors not detected as active), and the composite criterion $\log(\text{TPR} / \text{FPR}$) which summarizes the trade-off between maximizing true discoveries and minimizing false positives. We use 1{,}000 simulations unless otherwise indicated, and all simulations were run using the \verb|doParallel| package in R on 4 cores.

In all, the simulation included: the four effect sizes specified in the previous paragraph, ten combinations of design type and design size in Table \ref{tab:design_parms}, and the thirteen analysis methods in Table \ref{tab:analysis_methods} for a total of 520 simulation conditions.

\subsection{Simulation Results for Pool Construction Comparisons} \label{sec:design_results}

Using the $\log(\text{TPR}/\text{FPR})$ measure, we can average over the design sizes, analysis methods and effect sizes to see which design method performs the best. If we do that, CRowS is slightly preferred to MAPS (average 4.71 vs. 4.64). CRowS and Random appear to be equivalent for the cases that are fully balanced ($k=640$ and $k=1280$), but Random designs are unavailable for the unbalanced cases.

\begin{figure}
      \begin{subfigure}{\textwidth}
    \centering
    \includegraphics[width=0.9\textwidth]{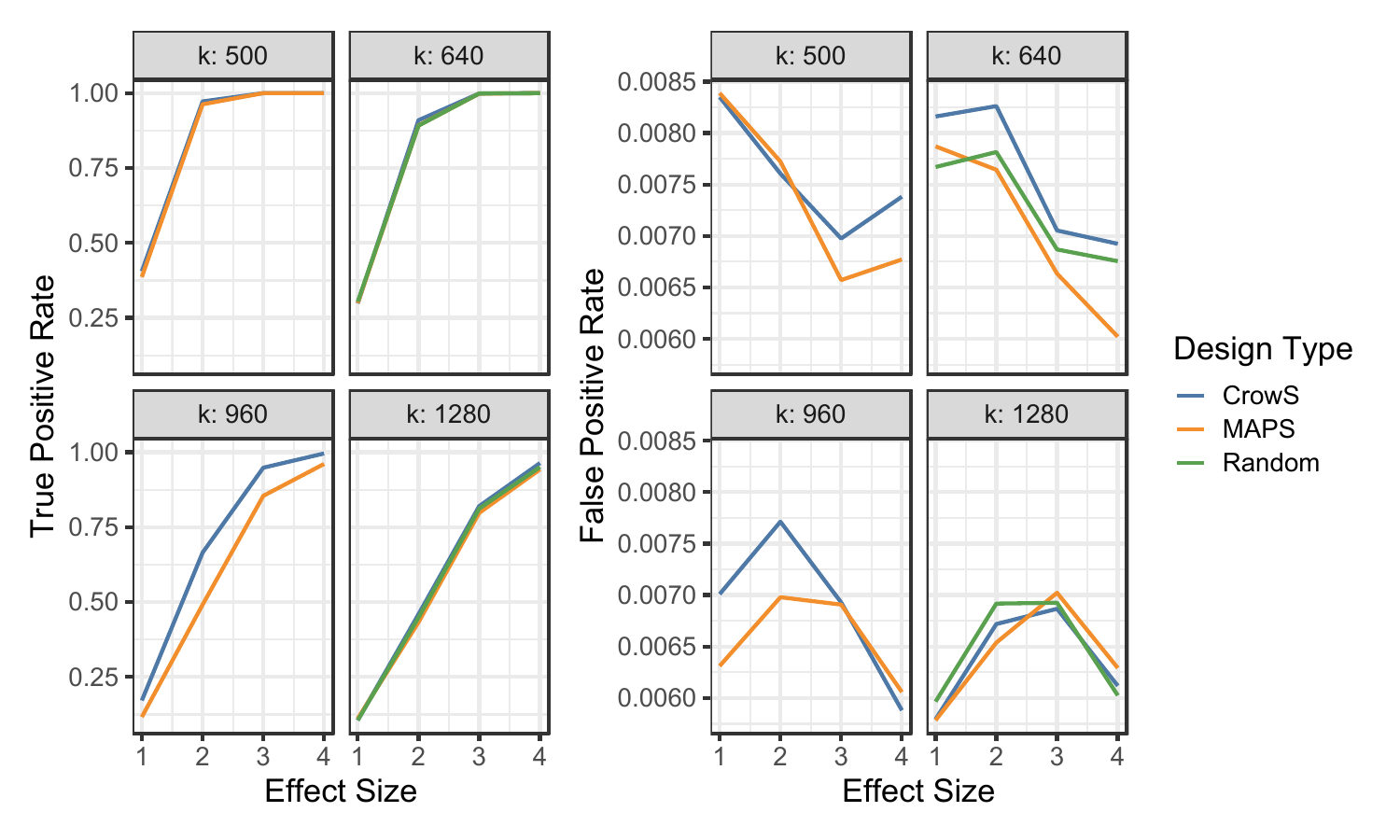}
    \caption{Plot of TPR and FPR, as a function of effect size and design size.}
  \end{subfigure}

  \vspace{1em}

    \begin{subfigure}{\textwidth}
    \centering
    \includegraphics[width=0.72\textwidth]{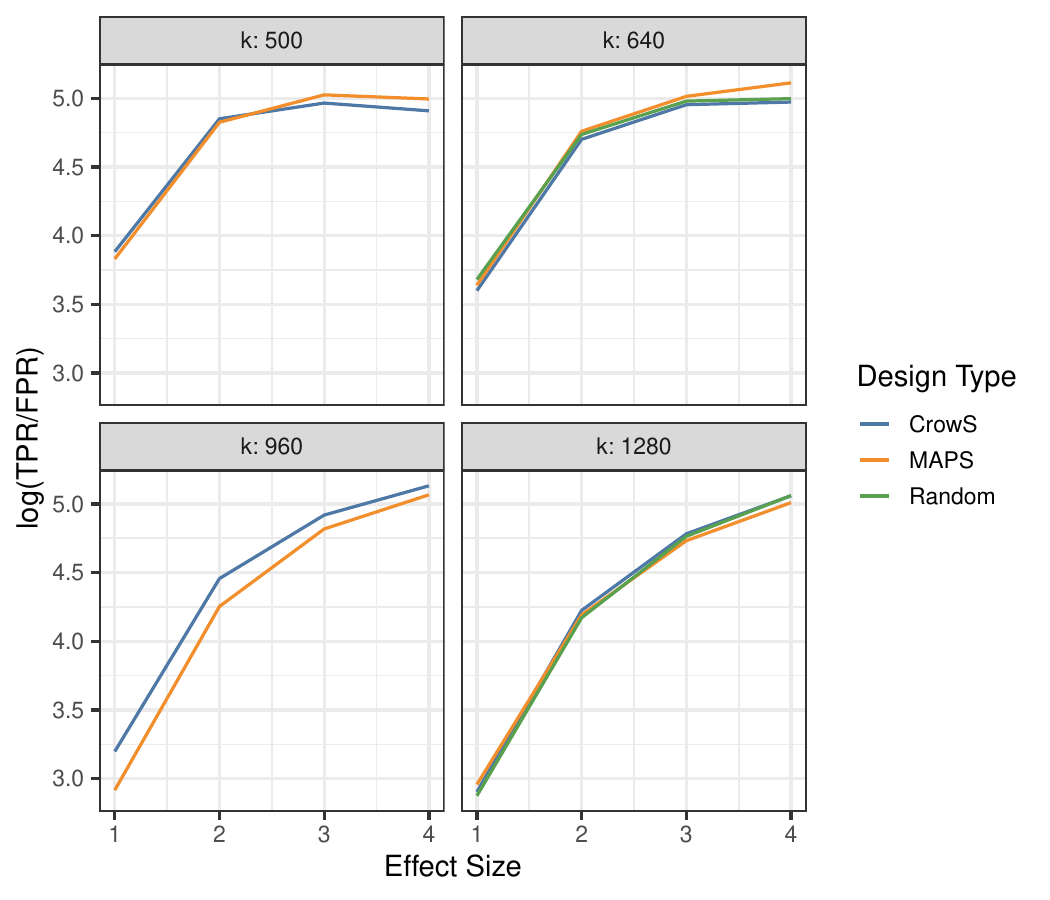}
    \caption{Plots of log(TPR/FPR), as a function of effect size and design size.}
  \end{subfigure}

  \caption{Comparison of three pool construction methods by simulation. In all cases, $\sigma=1$, there is $\lceil 0.01k \rceil$ active factors, and hits are detected using the Gauss-Lasso method of Section \ref{sec:methods} with threshold $\sigma/8$.}
  \label{fig:designs}
\end{figure}

To provide further insight, we present visualizations of the design methods. In Figure \ref{fig:designs}, for simplicity of presentation, we show results only for the Gauss-Lasso with threshold $\tau=\sigma/8$. We see that no design method dominates the others across the conditions considered. With the exception of the $k=960$ setting, performance across designs is similar. CRowS does show superiority to MAPS for the $k=960$ design, while the Random design with $nc=ka$ cannot even be constructed. 

Given its performance, flexibility and statistical grounding as a criterion, we generally recommend the CRowS designs to construct pooled high-throughput screens.

\subsection{Simulation Results for Comparisons of Analysis Methods} \label{sec:analysis_comparisons}

In this section we focus on the analysis of pooled HTS experiments, describing simulations to adjudicate between the statistical analysis methods described in Section \ref{sec:analysis_methods}, using the design sizes shown in Table \ref{tab:design_parms}. The basic goal is to understand whether the Gauss-Lasso---a method recommended for smaller, unconstrained supersaturated designs---is competitive or superior to other regularization methods like the nonnegative lasso and the Elastic Net. We also wish to learn whether our proposed method, which uses a $\lambda$-specific threshold taking advantage of our knowledge of effect directions, is an improvement over the other methods. Similar to the previous section, we will evaluate the methods using the $\log(\text{TPR}/\text{FPR})$ measure, and then provide some additional visualization to illustrate the results.

The last column in Table \ref{tab:analysis_methods} provides the average $\log(\text{TPR}/\text{FPR})$ for each of the 13 methods we considered. We note that for the thresholding-based methods, larger values of the threshold consistently yield better performance with respect to the $\log(\text{TPR}/\text{FPR})$ metric. The $\lambda$-specific Gauss-Lasso performs the best overall, with the $\tau_{\lambda}=\text{max}(\hat{\beta}_{\lambda})$ threshold preferred. 

\begin{figure}
      \begin{subfigure}{\textwidth}
    \centering
    \includegraphics[width=0.95\textwidth]{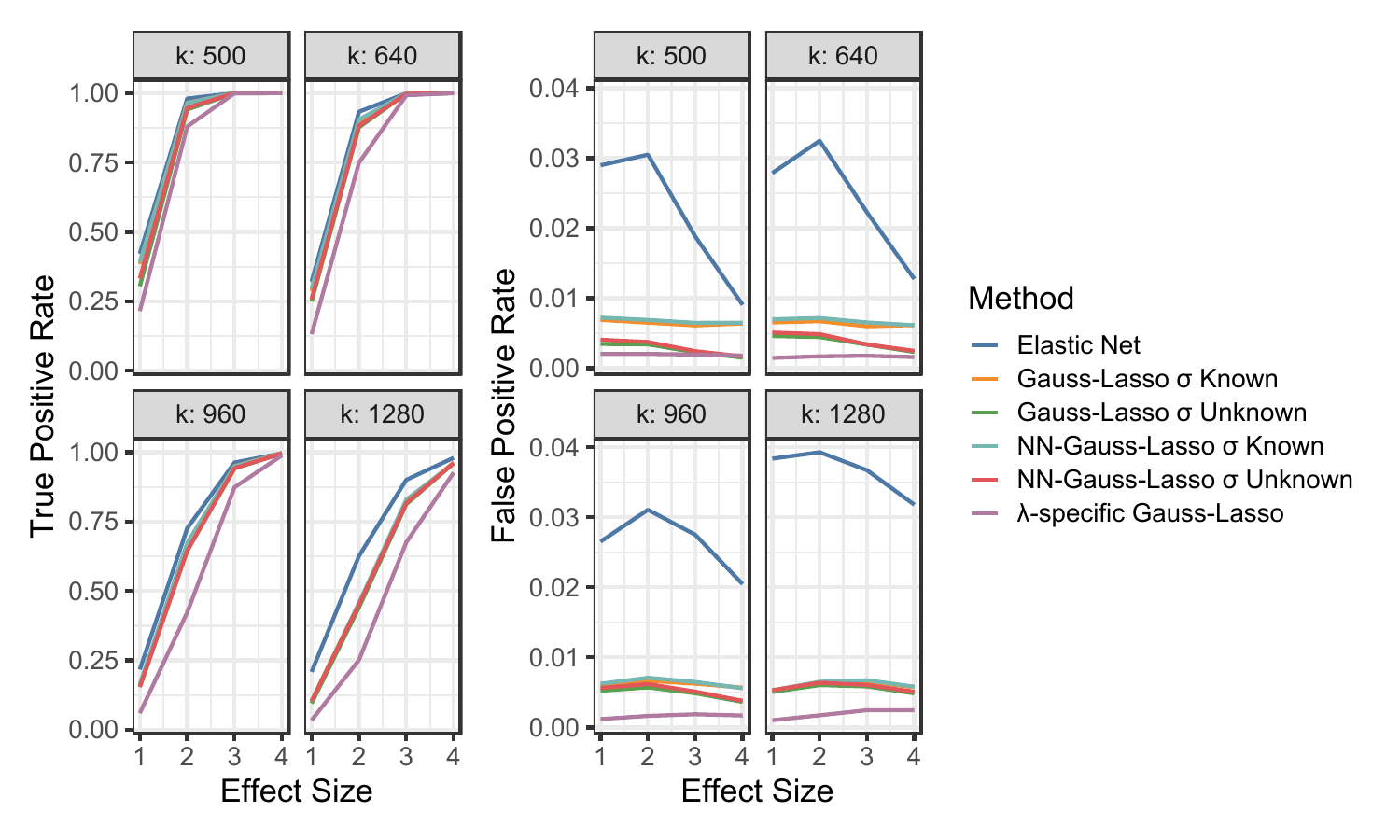}
    \caption{Plot of TPR (left) and FPR (right), as a function of effect size and analysis method. 
    }
  \end{subfigure}

  \vspace{1em}

    \begin{subfigure}{\textwidth}
    \centering
    \includegraphics[width=0.68\textwidth]{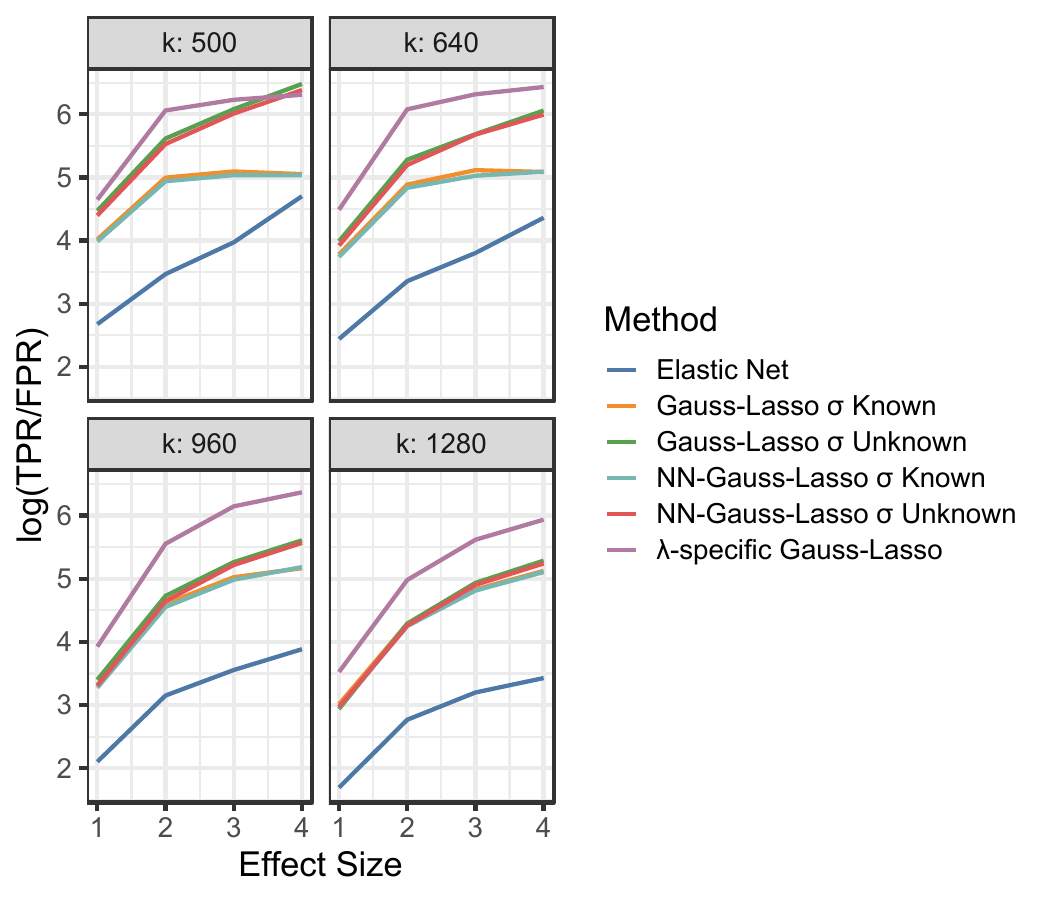}
    \caption{Plots of log(TPR/FPR), as a function of effect size and design size.
    }
  \end{subfigure}

  \caption{Comparison of a subset of the analysis methods in Table \ref{tab:analysis_methods}, for CRowS designs only. The methods are: the Elastic Net; Gauss-Lasso with $\tau=\sigma/4$ ($\sigma$ known); Gauss-Lasso with $\tau=0.5 \times \text{max}(|\hat{\beta}_{\lambda=0}|)$ ($\sigma$ unknown); Non-Negative Gauss-Lasso with $\tau=\sigma/4$ ($\sigma$ known); Non-Negative Gauss-Lasso with $\tau=0.5 \times \text{max}(|\hat{\beta}_{\lambda=0}|)$ ($\sigma$ unknown); and the $\lambda$-specific Gauss-Lasso with $\tau_{\lambda}=\text{max}(\hat{\beta}_{\lambda})$.}
  \label{fig:methods}
\end{figure}

To provide additional context for this conclusion, we choose the following six methods to compare visually: the Gauss-Lasso with $\tau=\sigma/4$; the Gauss-Lasso with $\tau=0.5 \times \text{max}(|\hat{\beta}_{\lambda=0}|)$; the $\lambda$-spcific Gauss-Lasso with $\tau_{\lambda}=\text{max}(\hat{\beta}_{\lambda})$; the non-negative Gauss-Lasso with $\tau=\sigma/4$; the non-negative Gauss-Lasso with $\tau=0.5 \times \text{max}(|\hat{\beta}_{\lambda=0}|)$; and the Elastic Net. Interestingly, we find in Figure \ref{fig:methods} that though the $\lambda$-specific Gauss-Lasso is best on the $\log(\text{TPR}/\text{FPR})$ measure, it is inferior with respect to TPR though better in terms of FPR than the other methods. In fact the Elastic Net method---though the worst on our metric---has the largest TPR among the methods because its FPR is much larger than the other approaches. Thus, the preference of one method over the other comes down to weighing the relative importance of detecting true hits (TPR) versus avoiding spurious effects (FPR). The Gauss-Lasso $\sigma$ Unknown (that is, the Gauss-Lasso with $\tau=0.5 \times \max(|\hat{\beta}_{\lambda=0}|)$) has a relatively high TPR and a relatively high $\log(\text{TPR}/\text{FPR})$, so it could be a good compromise between the Elastic Net and the $\lambda$-specific Gauss-Lasso. In this application, it is particularly important to reduce the number of false positives to reduce follow-up costs (see the next section), so though we recognize that the $\log(\text{TPR}/\text{FPR})$ measure can grow arbitrarily large as the $\text{FPR}$ approaches 0 while the influence of $\text{TPR}$ is limited because its value is capped at 1, it aligns well with the goals of our setting. 

In the analysis of the real experiments in the subsequent section, we focus on the use of the $\lambda$-specific Gauss-Lasso, along with a secondary criterion that we describe next.

\subsection{Secondary Criteria} \label{sec:secondary}

Despite the small false positive rates shown in our simulation, in the application of these methods to real data, in which the total number of compounds studied may be in the hundreds of thousands, we can use a secondary criterion to further focus the experimenter's attention on the most promising hits. This will necessarily reduce the true positive rate, but also the number of false positives.

The basic idea is to take the initial set of compounds identified by an analysis method as discussed in Sections \ref{sec:analysis_methods} and \ref{sec:analysis_comparisons}, and choose a subset of those whose pools are consistently and meaningfully inhibited, rather than compounds involved in only one or two inhibitory pools; that is, we want to focus on compounds whose wells are usually inhibitory at a level of practical significance. Assume a primary analysis method $P$ which produces a set of potentially active compounds $D_{P}$. Then, we propose to choose a subset $D_{S} \subseteq D_{P}$, where $D_{S}$ includes the set of compounds such that at least $(100 \times p_{s})$\% of the wells which include compound $j \in D_{S}$ are inhibitory. We define well $i$ to be inhibitory if $Y_{i}>\mu + r\sigma$, where $\mu$ is the average response for a well with no active compounds, $r$ is a user-specified value, and WLOG we assume positive effects are of interest. In our case, we considered $p_{s} \in \{0.75, 1\}$ and $r \in \{2,3\}$. Practically, for the $k=640$ CRowS design for which each compound appears in 4 wells, this means that we considered four secondary criteria: ``3 Wells $>$ 2 SD (above $\mu$)'', ``4 Wells $>$ 2 SD'', ``3 Wells $>$ 3 SD'', and ``4 Wells $>$ 3 SD''.

\begin{figure}[H]
    \centering
    \includegraphics[width=\linewidth]{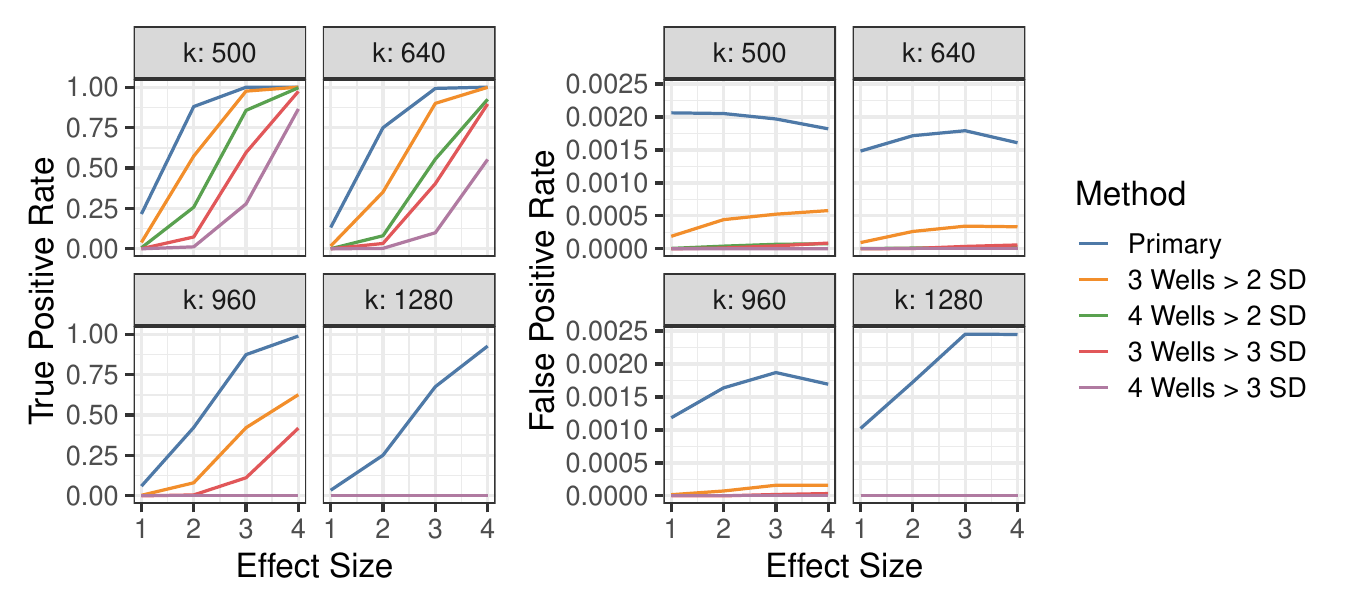}
    \caption{Comparison of FPR and TPR in CRowS designs for differing secondary criterion, using the $\lambda$-specific Gauss-Lasso with $\tau_{\lambda}=\text{max}(\hat{\beta}_{\lambda})$ primary thresholding criterion.}
    \label{fig:CRowS_secondary}
\end{figure}

We did a simulation of this secondary analysis strategy, using CRowS with $P \equiv \lambda$-specific Gauss-Lasso using $\tau_{\lambda} = \max(\hat{\beta}_{\lambda})$. The simulation settings were the same as with the previous simulations in Section \ref{sec:comparisons}; that is, with $k \in \{500, 640, 960, 1280\}$ and $\beta \in \{1,2,3,4\}$. We show TPR and FPR for the secondary criteria compared with the results of $P$ without a secondary criterion in Figure \ref{fig:CRowS_secondary}. As expected, both the TPR and FPR plummet as the secondary criteria get more strict. If we focus on $k=640$, which is the screen size we ended up using in the real experiments, we see that only the most strict ``4 Wells $>$ 3 SD'' secondary criterion have unacceptably low power for the largest effect. However, this highlights that these secondary criteria are really only viable if the experimenter is willing forgo the detection of small or even medium-sized effects. But the advantage is also important: the number of false positives is much more manageable. With a FPR of 0.0015 as in the analysis with no secondary criterion, investigators would be chasing 15 false positives for a 10{,}000 compound screen and 1{,}500 for a million-compound screen. But with the ``3 Wells $>$ 3 SD'' secondary criterion with an FPR of around $0.00006$, one would expect less than 1 and 60 false positives, respectively.

\section{Designing and Analyzing the Anti-Microbial Screens} \label{sec:application}

In this section, we describe the process by which our screens were designed, executed, validated, and used to identify promising compounds for the MtlD system. First, we conducted several proof-of-concept screens to ensure that our methods could detect known hits, for different levels of screening boldness. Then, we report results of a small screening campaign the goal of which was to identify new compounds that inhibit MtlD. Recalling Section \ref{sec:application_description}, each plate must be measured using a MUT and WT assay in order to find a drug that inhibits the WT but not the MUT.

We used the basic methodology described in \citep{smucker2025large} to construct CRowS designs, then used the Lasso to obtain lists of potential hits. Details of our analysis are provided in the results below, but ultimately we used the $\lambda$-specific Gauss-Lasso along with the secondary criteria of Section \ref{sec:secondary} to construct hit lists. Note that we report on actual experiments conducted before we did the extensive testing described in Sections \ref{sec:methods} and \ref{sec:comparisons}. Thus, our recommendations from the comparative work may not map precisely onto what we actually describe in this section. The purpose of reporting these results is to demonstrate a high-quality solution to the screening problem presented in Section \ref{sec:application_description}, while providing insight and guidance for other HTS applications.

\subsection{Proof-of-Concept Screening Results} \label{sec:pilot_results}

A first step when considering pooling for a particular assay and for the type of compounds to be screened is to ensure that the pooling method can recognize known hits. This is important because it is possible for pooled HTS to fail. If there are large and/or numerous interactions, the ability to identify main effects may be degraded \citep[see results referenced in the Discussion of][]{smucker2025large}. There have also been reports of compounds reacting together in wells in therapeutically uninteresting ways \citep[e.g.,][]{feng2006synergy}. Thus, for the proof-of-concept screens described in this section, we spiked the screens with a known pseudohit (Carbenicillin, a drug known to hit both the WT and MUT assays, denoted \verb$CntP1A03$), as well as a known true hit (Chloramphenicol, a drug known to hit WT but not MUT, denoted \verb$CntP1A04$). We used compounds from the MedChem Express Diversity library (5{,}000 Scaffold Library; Cat. No.: HY-L902) We also wished to investigate the extent to which our screening method could be successfully stretched. This led us to executing the following pooled HTS experiments, using CRowS designs:
\begin{itemize}
    \item $(n=320, k=500, c=8)$,  with $a_{\text{min}}=5$ and $a_{\text{max}}=6$. That is, each compound appears in either 5 or 6 pools.
    \item $(n=320, k=640, c=8)$, with $a=4$.
    \item $(n=320, k=960, c=8)$, with $a_{\text{min}}=2$ and $a_{\text{max}}=3$.
    \item $(n=320, k=1280, c=8)$, with $a=2$.
\end{itemize}

\begin{figure}[H]
    \centering
    \includegraphics[width=\linewidth]{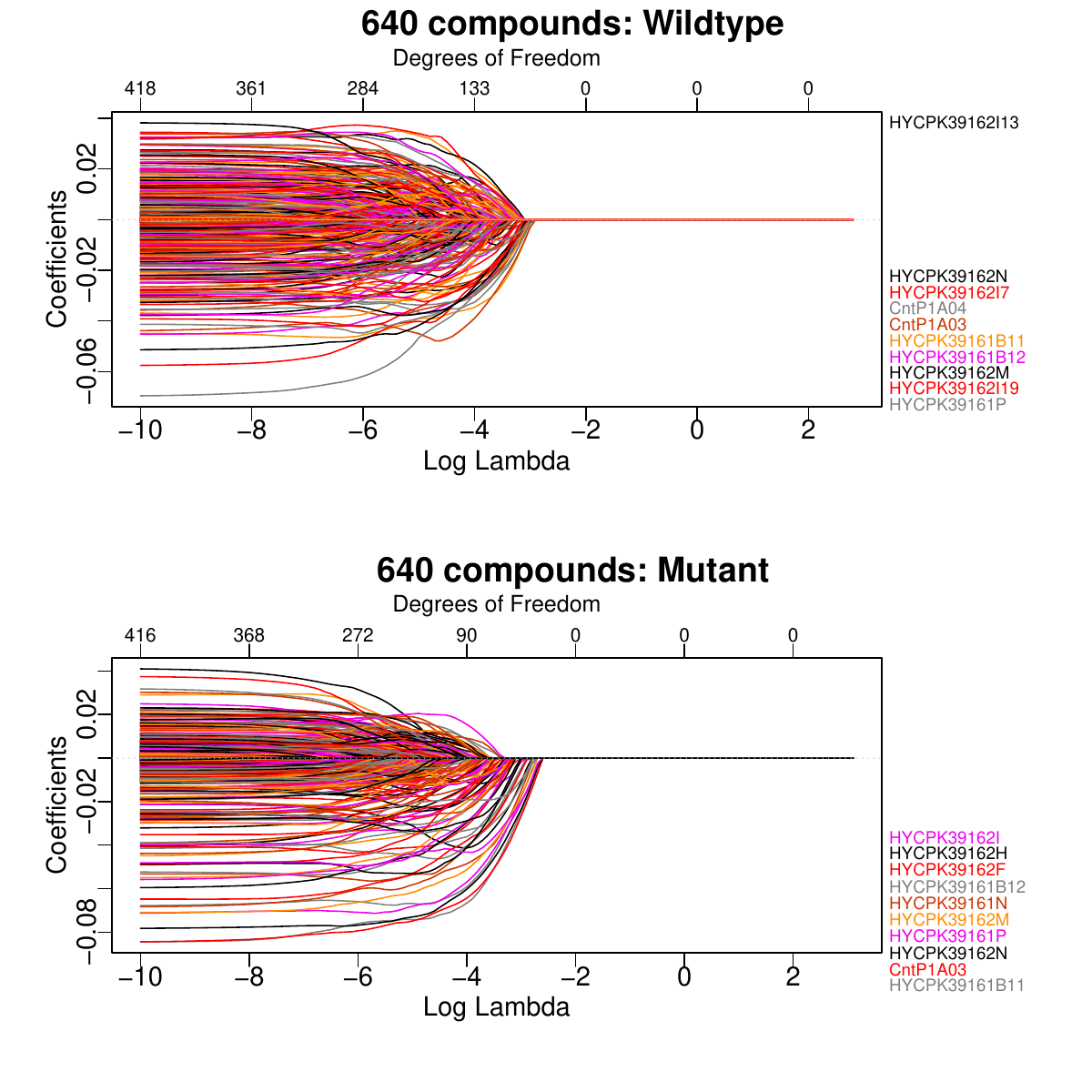}
    \caption{Lasso profile plots for $(n=320, k=640, c=8)$ proof-of-concept CRowS design. The plot annotates the top 10 compounds in terms of magnitude at the smallest $\lambda$.}
    \label{fig:CRowS640PoC}
\end{figure}

\begin{figure}[H]
    \centering
    \includegraphics[width=\linewidth]{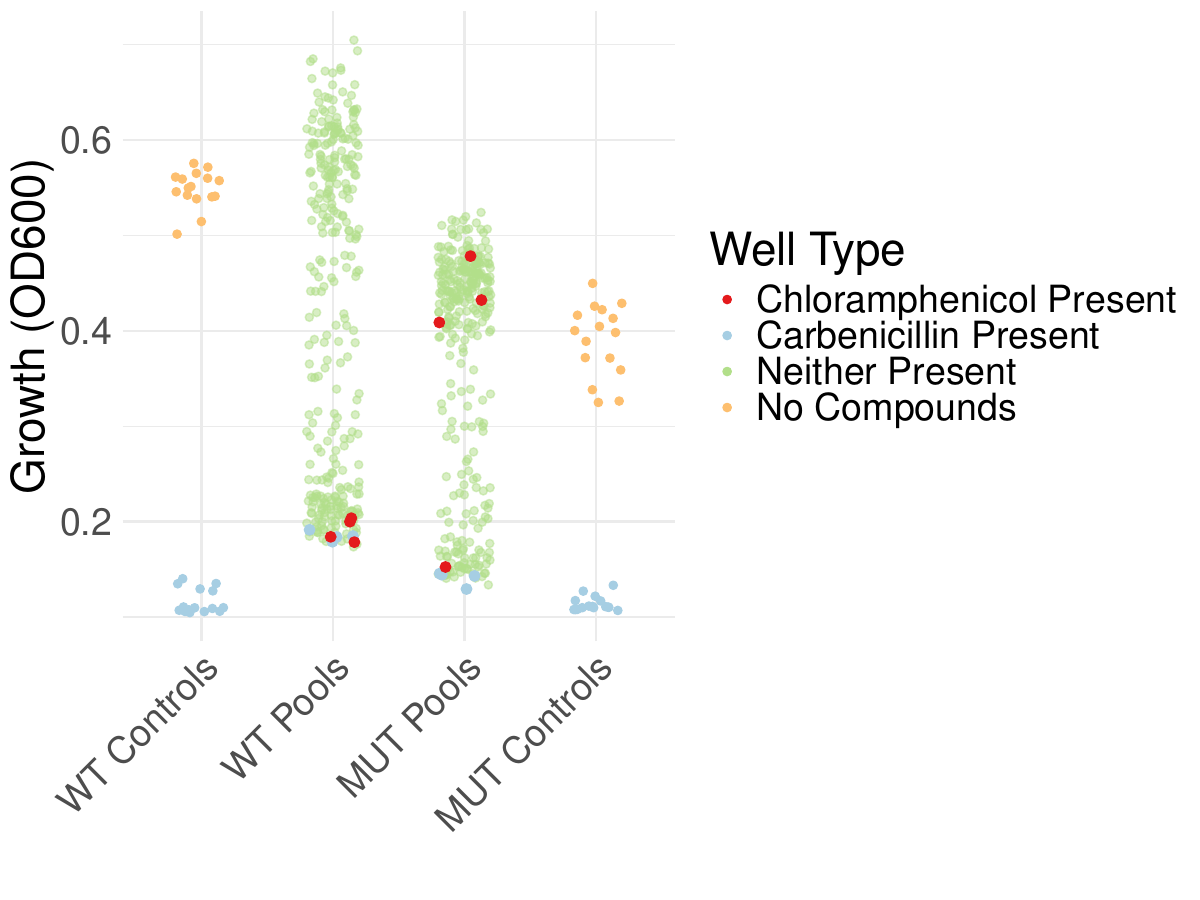}
    \caption{Plots of the data for $(n=320, k=640, c=8)$ proof-of-concept CRowS design. The control wells consist of Carbenicillin as a positive control and DMSO (no compounds) as a negative control.}
    \label{fig:CRowS640PoC-jitter}
\end{figure}

The results of these experiments were promising. Examining the Lasso profile plots, along with the data broken out by the spiked compounds of interest, showed that pooling was working in both the $k=500$ and $k=640$ settings. Figures \ref{fig:CRowS640PoC} and  \ref{fig:CRowS640PoC-jitter} show these plots for $(n=320, k=640, c=8)$. In Figure \ref{fig:CRowS640PoC} observe that \verb$CntP1A03$ is annotated as one of the most prominent among the profiles in both the WT and MUT assay, while \verb$CntP1A04$ only appears prominent in the WT. This suggests, correctly, that \verb$CntP1A03$ is a pseudohit and that \verb$CntP1A04$ scores as a true inhibitor. This is visually confirmed in Figure \ref{fig:CRowS640PoC-jitter}, where \verb$CntP1A03$ clearly inhibits both WT and MUT, but \verb$CntP1A04$ inhibits only WT. In order to reduce the subjectivity in these judgments, we used the $\gamma$-specific Gauss-Lasso described in Section \ref{sec:analysis_comparisons}, with $r=0.9$. Using this procedure, we found three compounds that hit WT but not MUT, including \verb$ControlP1A04$, as expected. The two false positives are consistent with a false positive rate of about 0.3\%. We found similar results for $(n=320, k=500, c=8)$, with the profile plots, data, and $\gamma$-specific Gauss-Lasso all identifying \verb$ControlP1A04$ as an inhibitor, and the latter procedure identifying in addition four presumed false positives.

The results from the more aggressive $(n=320, k=960, c=8)$ and $(n=320, k=1280, c=8)$ experiments were also promising, but more ambiguous. Visually, the profile plots yielded no hint of the expected pseudohit and true hit, and the $\gamma$-specific Gauss-Lasso failed to identify the compounds. However, plotting the compounds separately from the rest of the data for the experiments still suggested that \verb$CntP1A03$ inhibited both, while \verb$CntP1A04$ inhibited just the WT. Still, with just 2 or 3 pools per compound, there wasn't enough statistical information to clearly identify them. We have included the plots for the $(n=320, k=1280, c=8)$ experiment in the Supplementary document. Based on these results, we moved forward to the screening stage with plans to use the $(n=320, k=640, c=8)$ design.

\subsection{Screening Results} \label{sec:screening_results}

Following the proof-of-concept experiments, we conducted a small-scale screen consisting of 16 plates, called PLINGS. The compounds used in this screen were from ChemBridge Macrocycle Library (10{,}000 Macrocycles; 2018 version; N1558-1) As per the test screens, each PLING included 640 unique compounds, which means the total number of compounds represented in this initial screen is 10{,}240. For this system, the expectation is that true hits may be as rare as 1 in 100{,}000. Thus, there is no certainty that in this first screen we will find a true hit. Still, we present the results that we have obtained to this point.

In the Supplementary Document, we have included the boxplots from the PLINGS. Predictably, with real data there are ambiguities. For instance, it is clear that in some plates there are a relatively small number of wells that demonstrate inhibition, and in others there are indications of potential activation. Also, some plates have more apparently inhibited wells than would be expected in such a sparse system. Finally, the MUT vs. WT were expected to have similar distributions, but in reality the MUT assay is consistently centered lower than the WT assay (This is certainly due to the WT utilizing mannitol in the medium while the MUT cannot. This could be remedied in future assays by reducing the concentration of mannitol). 

\begin{figure}[htbp]
  \centering

    \begin{subfigure}{\textwidth}
    \centering
    \includegraphics[width=0.72\textwidth]{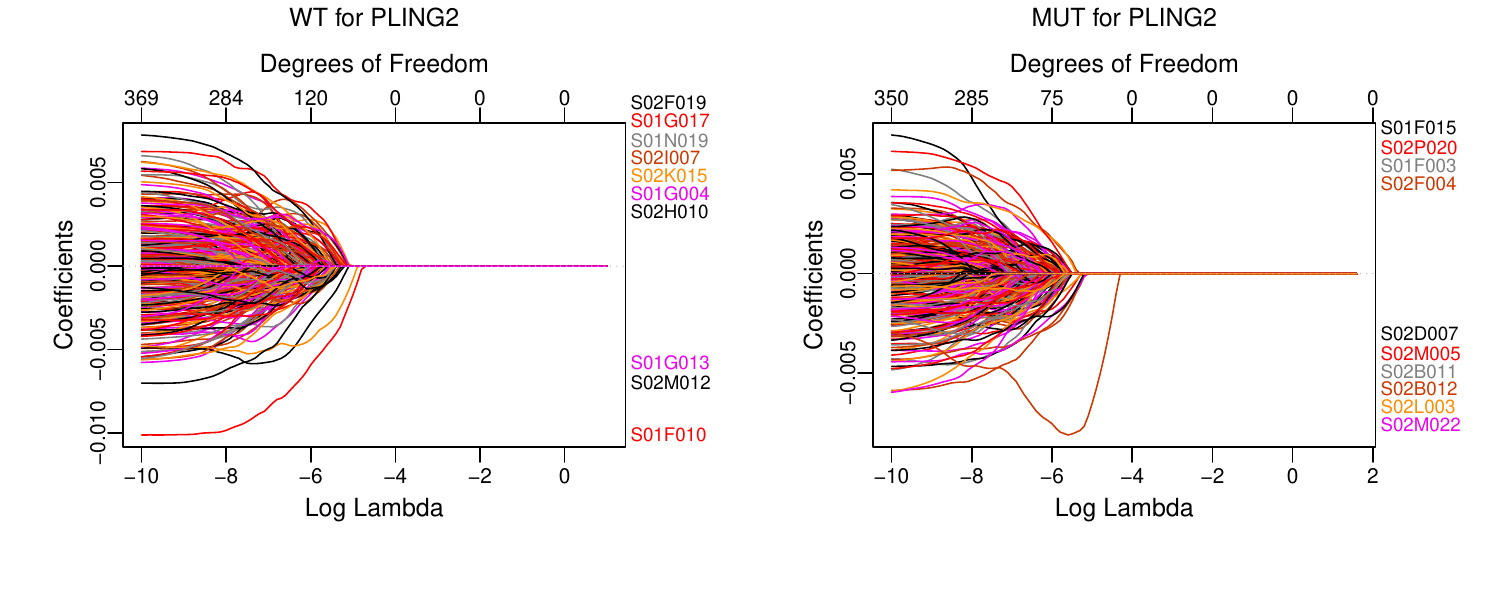}
    \caption{\phantom{}}
  \end{subfigure}

  \vspace{1em}

    \begin{subfigure}{\textwidth}
    \centering
    \includegraphics[width=0.72\textwidth]{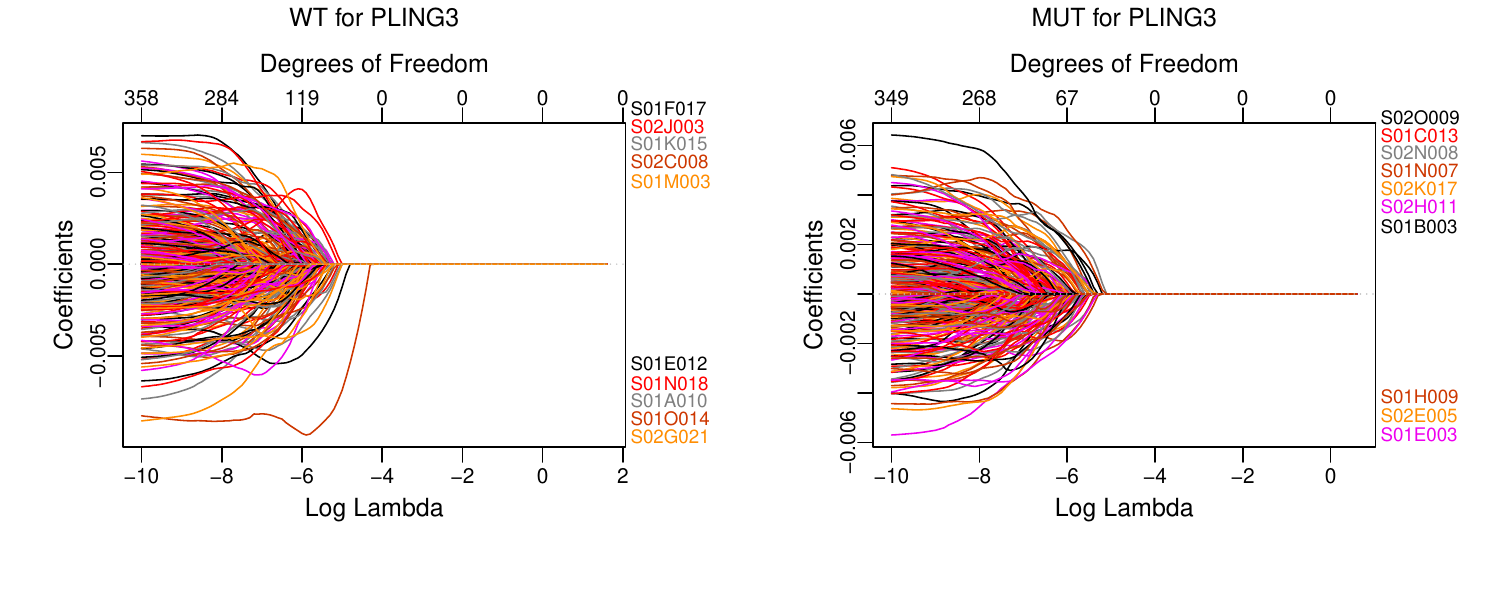}
    \caption{\phantom{}}
  \end{subfigure}

  \vspace{1em}

      \begin{subfigure}{\textwidth}
    \centering
    \includegraphics[width=0.72\textwidth]{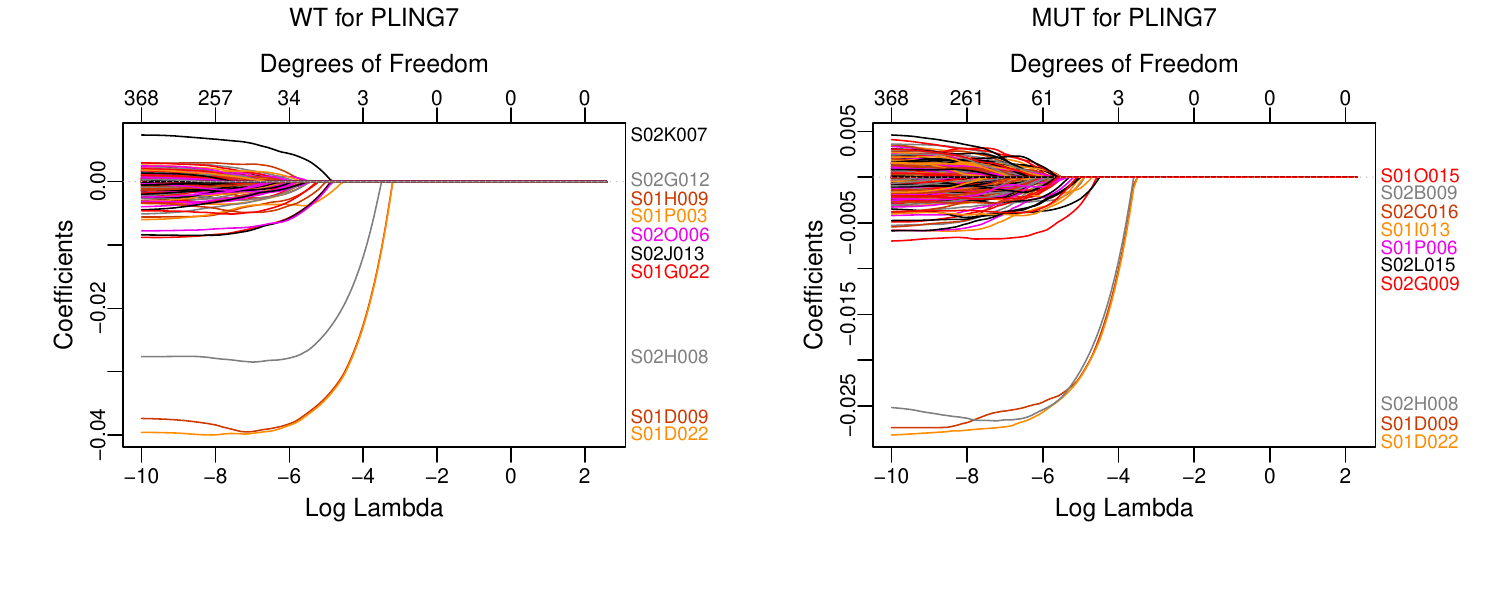}
    \caption{\phantom{}}
  \end{subfigure}

  \vspace{1em}

    \begin{subfigure}{\textwidth}
    \centering
    \includegraphics[width=0.72\textwidth]{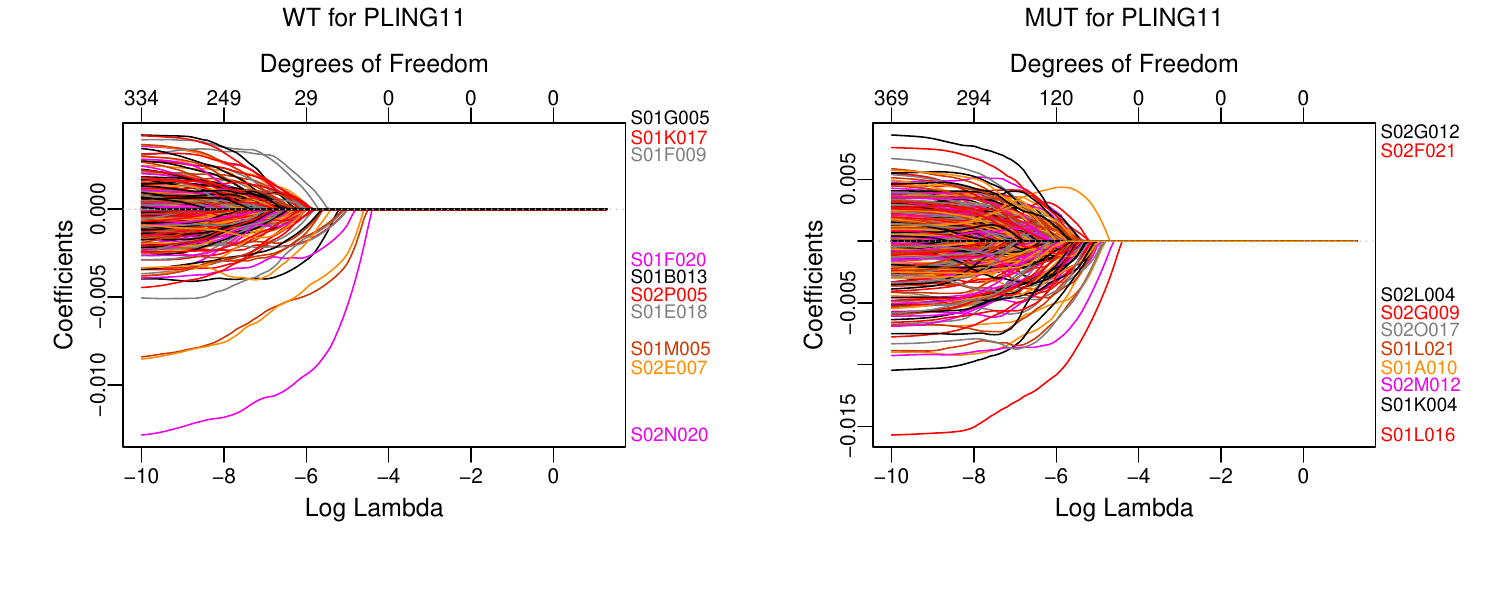}
    \caption{\phantom{}}
  \end{subfigure}

  \caption{Four profile plots.}
  \label{fig:profile-plots}
\end{figure}

Despite these data challenges, each plate was analyzed visually via profile plots. Though we don't include the profile plots from all of the PLINGs, in Figure \ref{fig:profile-plots} we provide four: (a) In PLING2 there is a promising compound \verb$S01F010$ that appears to inhibit in WT but not in MUT; (b) in PLING3 there are no clear inhibitors for either assay; (c) in PLING7, there appear to be three pseudohits, \verb$S02H008$, \verb$S01D009$, and \verb$S01D022$, because they hit on both assays; and (d) in PLING11 there are several possible true hits, \verb$S02N020$, \verb$S02E007$, and \verb$S01M005$, which show up on WT but not MUT.

The preceding analysis is relatively subjective, so we use the $\lambda$-specific Gauss-Lasso with $r=0.9$ to provide an objective hit list. To review, this Gauss-Lasso procedure sets any Lasso estimate to 0 if the estimate is greater than $-0.9\times \max{\hat{\beta}_{\lambda}}$. That is, the only estimates that survive are the ones that are relatively large and negative. This is done for both the WT and MUT assays, for each of the PLINGs separately. Then, for each PLING, we take an initial hitlist to be those that hit on WT but not on MUT. Using the 90\% $\lambda$-specific Gauss-Lasso yields 1.1\% of the studied compounds as potential hits. Though this seems to be a reasonable hit rate---even if most of them are false positives---but even in this small initial screen there are more than 100 hits identified, and if scaled to hundreds of thousands of compounds such a hit rate would place a tremendous burden on secondary screens.

In order to further screen the initial hitlists, we search among the potential hit compounds for those whose pools consistently exhibited substantial inhibition, using the method described in Section \ref{sec:secondary}. For the $(n=320, k=640, c=8)$ design, each compound appears in four wells, so we looked for promising compounds with three or four pools more than 2 or 3 standard deviations away from the median of the pools not containing the compound under consideration. To estimate the standard deviation, we used the robust estimator of $1.48 \times MAD$, where $MAD$ is the median absolute deviation of the remaining points. This was done in case other hit compounds exist in the remaining pools.

\begin{figure}[htbp]
  \centering

    \begin{subfigure}{\textwidth}
    \centering
    \includegraphics[width=0.72\textwidth]{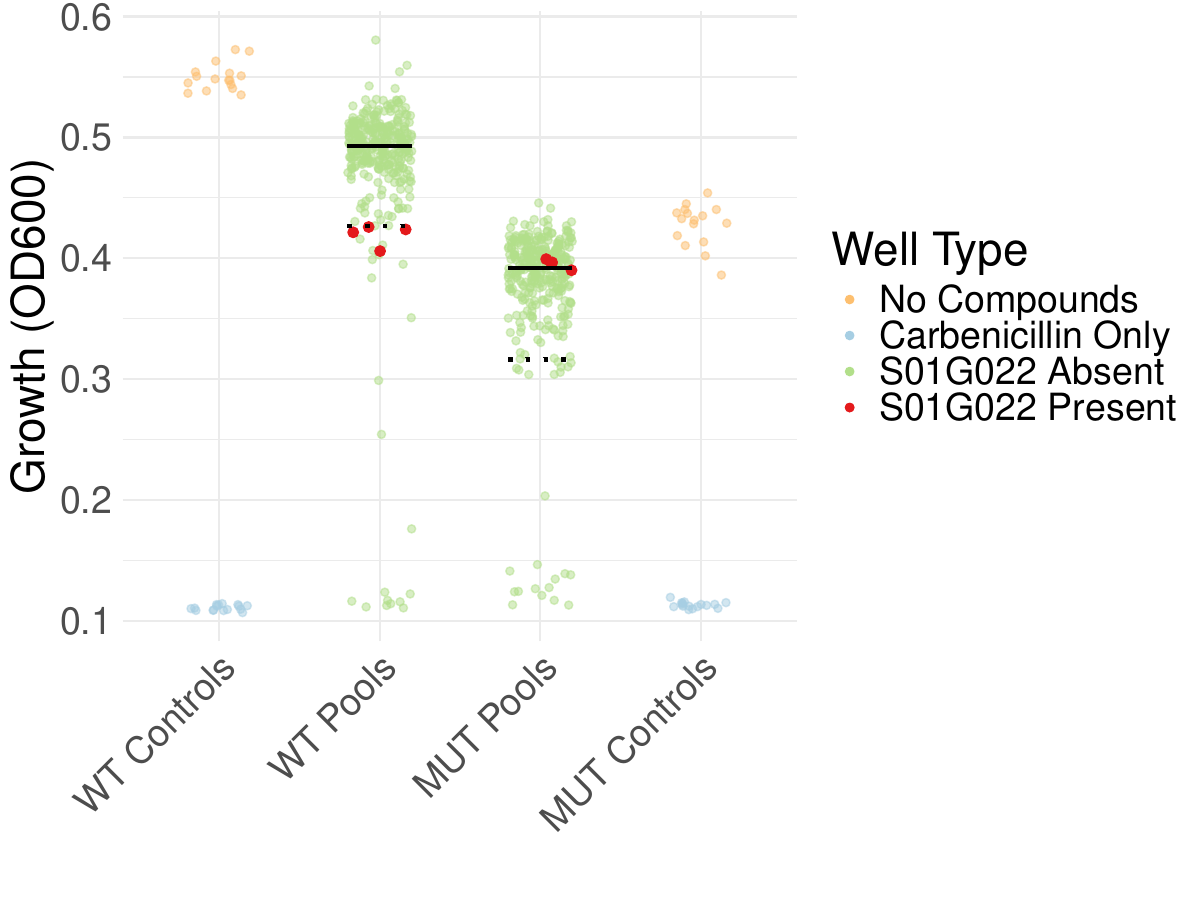}
    \caption{Promising compound \texttt{PLING7-S01G022} with all four of its pools beyond two standard deviations from the median for WT assay, while not inhibiting the MUT assay. The solid line for the WT pools is an estimate of the median; the dashed line is an estimate of 2 SD below the median.}
  \end{subfigure}

  \vspace{1em}

    \begin{subfigure}{\textwidth}
    \centering
    \includegraphics[width=0.72\textwidth]{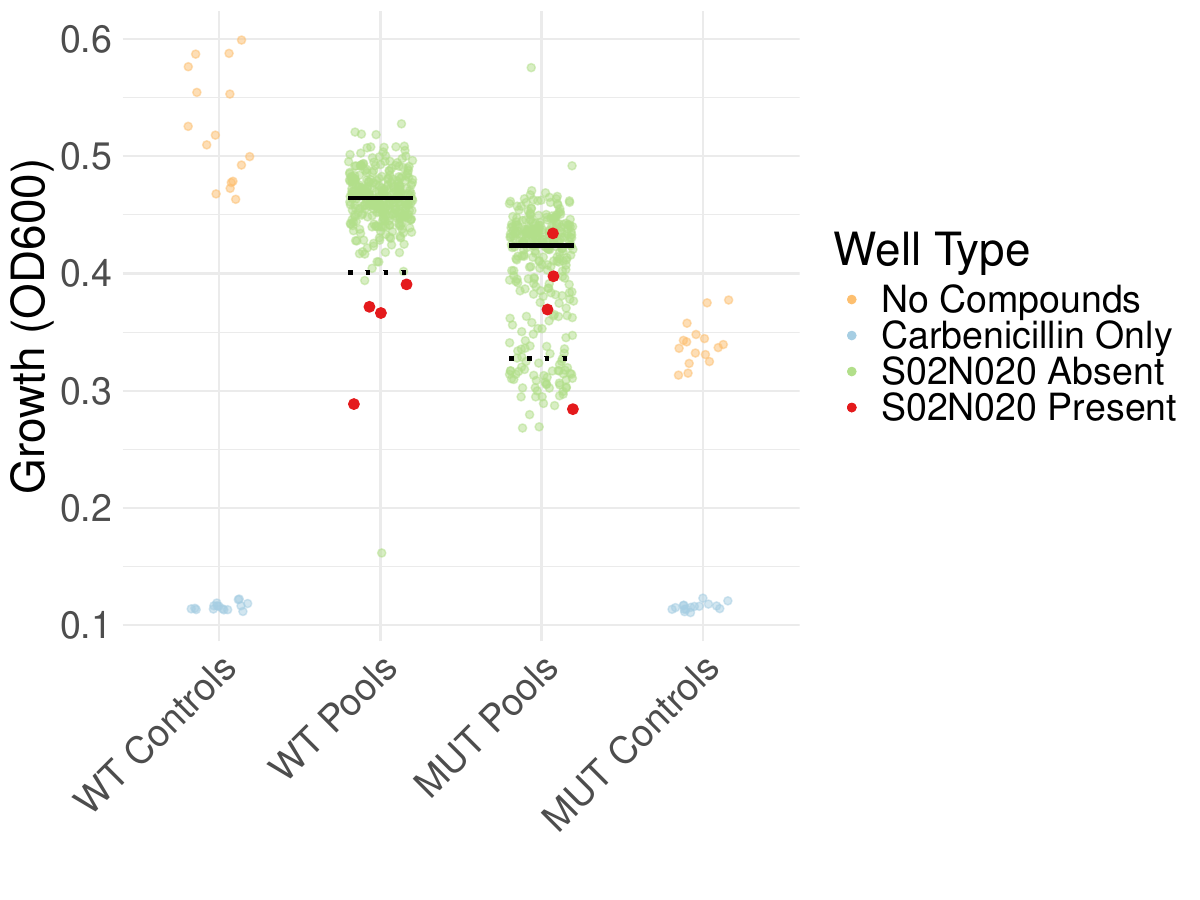}
    \caption{Promising compound \texttt{PLING11-S02N020} with all four of its pools beyond two standard deviations from the median for WT assay, while not obviously inhibiting the MUT assay. The solid line for the WT pools is an estimate of the median; the dashed line is an estimate of 2 SD below the median.}
  \end{subfigure}

  \caption{Two promising 2-SD compounds.}
  \label{fig:promising_compounds2}
\end{figure}

\begin{figure}
      \begin{subfigure}{\textwidth}
    \centering
    \includegraphics[width=0.72\textwidth]{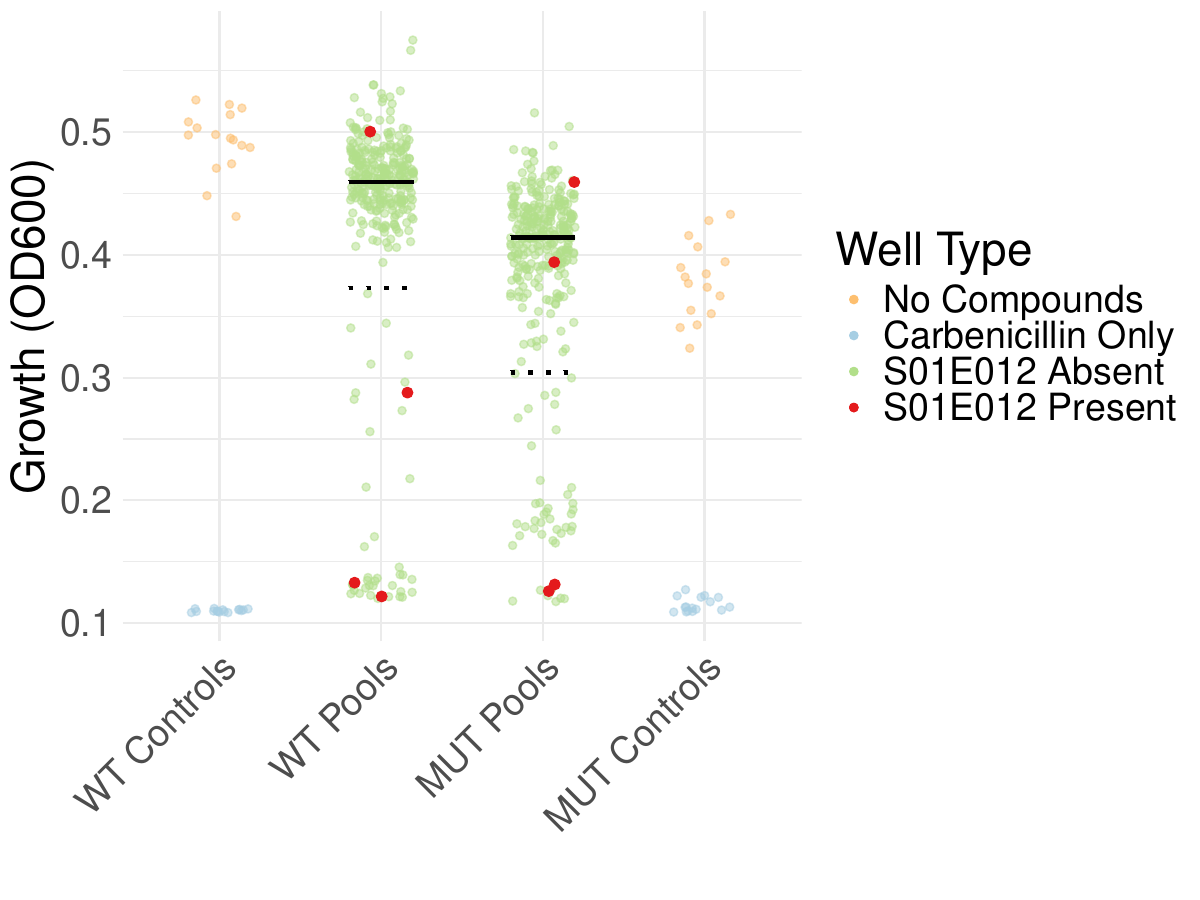}
    \caption{Promising compound \texttt{PLING5-S01E012} with three of its four pools beyond three standard deviations from the median, though is appears that two of its pools also inhibit the MUT assay. The solid line for WT is an estimate of the median; the dashed line is an estimate of 3 SD below the median.}
  \end{subfigure}

  \vspace{1em}

    \begin{subfigure}{\textwidth}
    \centering
    \includegraphics[width=0.72\textwidth]{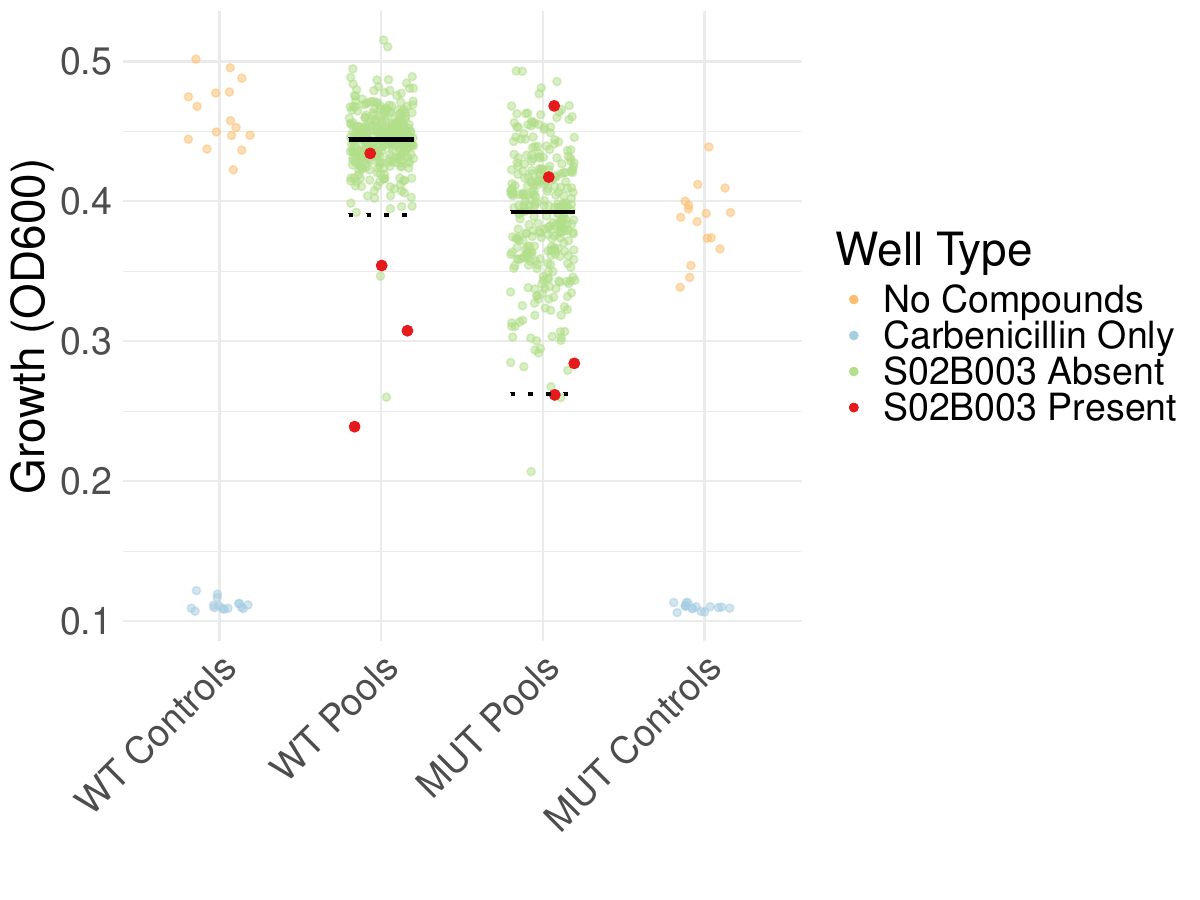}
    \caption{Promising compound \texttt{PLING6-S02B003} with three of its four pools beyond three standard deviations from the median of the WT assay. The solid line is an estimate of the median for WT; the dashed line is an estimate of 3 SD below the median.}
  \end{subfigure}

  \caption{Two promising 3-SD compounds.}
  \label{fig:promising_compounds3}
\end{figure}

Table \ref{tab:secondary_nums} identifies 8 unique compounds, arbitrarily named, that meet the 2 SD secondary criteria, which is that at least three of its four pools inhibit by 2 or more SDs. It also includes, naturally, the two compounds that meet the 3 SD secondary criteria. The secondary criteria eliminates the issue of compounds chosen by the method due to one or two outliers. Of these 8, we show in Figures \ref{fig:promising_compounds2} and \ref{fig:promising_compounds3} two compounds for which all four wells are 2 SD hits, as well as the two additional compounds which have three of four wells beyond 3 standard deviations. This methodology allows the researchers to visually focus on the very most promising compounds.

\begin{table}[ht]
    \centering
    \begin{tabular}{c|l||c|c}
    PLING & Compound & Number Beyond 2 SDs & Number Beyond 3 SDs \\
      \hline \hline
    1 & S01I013 & 3 & 0 \\
    5 & S01E012 & 3 & 3 \\
    5 & S01L010 & 3 & 2 \\
    6 & S02B003 & 3 & 3 \\
    7 & S01G022 & 4 & 0 \\
    7 & S02O006 & 3 & 0 \\
    8 & S01K008 & 3 & 2 \\
    11 & S02N020 & 4 & 2 \\
      \hline \hline
    \end{tabular}
    \caption{For the initial screen, the eight compounds that met the 2 SD secondary criterion, including the two compounds that met the 3 SD secondary criterion. In this screen, each compound appears in four pools.}
    \label{tab:secondary_nums}
\end{table}

So far, none of the promising compounds have been validated as true hits, but this is not unexpected in a system anticipated to be as sparse as this.

\section{Discussion} \label{sec:discussion}

In this work, one major contribution is to provide practitioners of pooling in HTS guidance regarding pool construction methods and screen analysis methods. Along the way, we introduce two improvements to previously-described analysis methods that align with our goal of reducing false positives. The second contribution of our work is to apply these methods to a real problem in search of a solution. This problem is the resistance to antibiotics that has progressed and a possible solution is the inhibition of an enzyme that neutralizes the toxic accumulation of the compound mannitol-1-phosphate in many bacteria.  We show how the designs and analysis methods we studied can be used in practice, with both an extensive pilot study---in which controls are included---and a small-scale screening campaign. Such an extensive set of experiments provides insights into the challenges and compromises that must be faced in real world data.

Regarding the comparisons of design methods, we compared several recent pool construction strategies and concluded that the CRowS approach of \citep{smucker2025large} offers a good combination of effectiveness and flexibility. We note that when the design parameters allow balance in terms of pool sizes and compound replication, it appears that all three methods we compared perform similarly. Of course, we are limited in the generality of our conclusion by the specificity of our simulation scenarios. But we would tentatively state that for such balanced cases, there is likely little difference between CRowS, MAPS, and the randomly constructed pools. In fact, we conjecture that for those balanced cases, ``random'' pools are sufficiently constrained to yield CRowS designs, though we have not proved this.

We also compared a number of methods to analyze pooled HTS data and concluded that for the present application---in which it is critical to reduce the number of false positives to near zero---a new $\lambda$-specific Gauss-Lasso method provides the best balance of TPR and FPR. This method takes advantage of the fact that we know that any true effect will be inhibitory. We also described an additional, secondary analysis method that further reduces the number of false positives. For the analysis method comparisons, the preferred method strongly depends on the balance the experimenter wishes to strike between true and false positive rate.  The $\lambda$-specific Gauss-Lasso, with its relatively low FPR but also lower TPR, aligns with the need of our application. But if it is important to identify more and smaller effects, the Elastic Net or the Gauss-Lasso with $\tau=0.5 \times \text{max}(|\hat{\beta}_{\lambda=0}|)$.

A critical assumption in pooling is that statistical interactions between compounds are not numerous nor large. \citep{smucker2025large} briefly investigated this and showed that interactions can have a large impact on whether hits are detected. There is also concern regarding promiscuous aggregation \citep{feng2006synergy} of compounds, in which molecules combine in unpredictable and unhelpful ways and prevent clean deconvolution of individual compound effects. It is beyond the scope of this work to investigate these things further, but the present work has demonstrated at least that strong hits are detectable in real-world, sparse settings such as the one investigated herein. 

\setcounter{figure}{0}
\renewcommand{\thefigure}{S\arabic{figure}}

\subsection*{Supplementary Material}

Supplementary Material consists of:
\begin{enumerate}[label=\Alph*]
    \item Supplementary document referred to in the main document. [Available at the end of this document]
    \item Pooled and control data from the pilot experiments of Section \ref{sec:pilot_results}, along with pooled and control data from the screening experiments of \ref{sec:screening_results}. [Unavailable in arXiv version]
\end{enumerate}

\subsection*{Acknowledgements}

ChatGPT/Co-Pilot was used to assist in writing analysis, simulation and/or figure-generation code. We thank the Drug Discovery Shared Resource High Throughput Screening Lab at The Ohio State University Comprehensive Cancer Center for technical support.

\subsection*{Funding}

BRP, MW, and ZL thank the OSU Comprehensive Cancer Center (2P30 CA016058) for financial support. MW acknowledges the NIH support of R50 CA243786. ER was supported by PHS grant NIH T32 AI165391.

\subsection*{Conflicts of Interest}

The authors declare no conflicts of interest.

\subsection*{Data Availability Statement}

The pooled data from Section 5 is available in the Supplementary Material. Raw data, control data, and/or codes for data processing and simulations are available upon request of the authors.

\bibliographystyle{unsrt} 
\bibliography{main.bib}

\section*{Supplementary Document}

\subsection*{CRowS vs. Orthogonal Pooling}\label{sec:OP}

\begin{figure}[H]
    \centering
    \includegraphics[width=\linewidth]{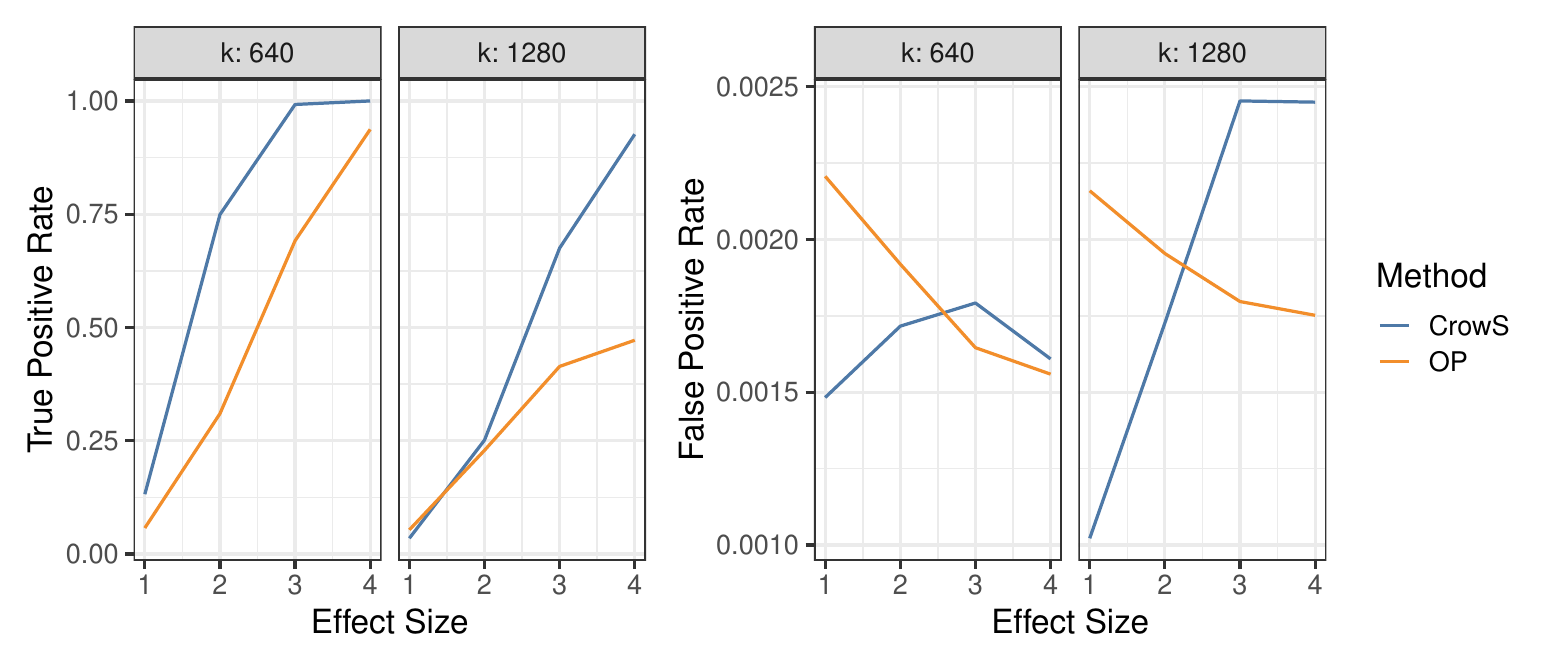}
    \caption{Comparison of FPR and TPR between CRowS and Orthogonal Pooling designs using the $\lambda$-specific Gauss-Lasso with $\tau_{\lambda}=\text{max}(\hat{\beta}_{\lambda})$ primary thresholding criterion for CRowS.}
    \label{fig:OP_results}
\end{figure}

Here we report results of CRowS vs. Orthogonal Pooling (Figure \ref{fig:OP_results}, where CRowS is analyzed using the Gauss-Lasso with $\tau_{\lambda}=\text{max}(\hat{\beta}_{\lambda})$ while the Orthogonal Pooling design is analyzed using a traditional, non-statistical modeling method described below.

For orthogonal pooling designs, the data was generated in a manner identical to our primary simulations, with effect sizes $\beta = (1,2,3,4)$ and error $\sigma^2 = 1$. The simulations were carried out with 10,000 iterations, and, at each iteration, a compound was designated a hit if the values of both observed wells ($y_i, y_j$) in which the compound was present were greater than the $95^{th}$ percentile ($y_{0.95}$) of all observed well values.

\subsection*{Proof-of-concept Screening Results for $(n=320, k=1280, c=8)$}

\begin{figure}[h]
    \centering
    \includegraphics[width=\linewidth]{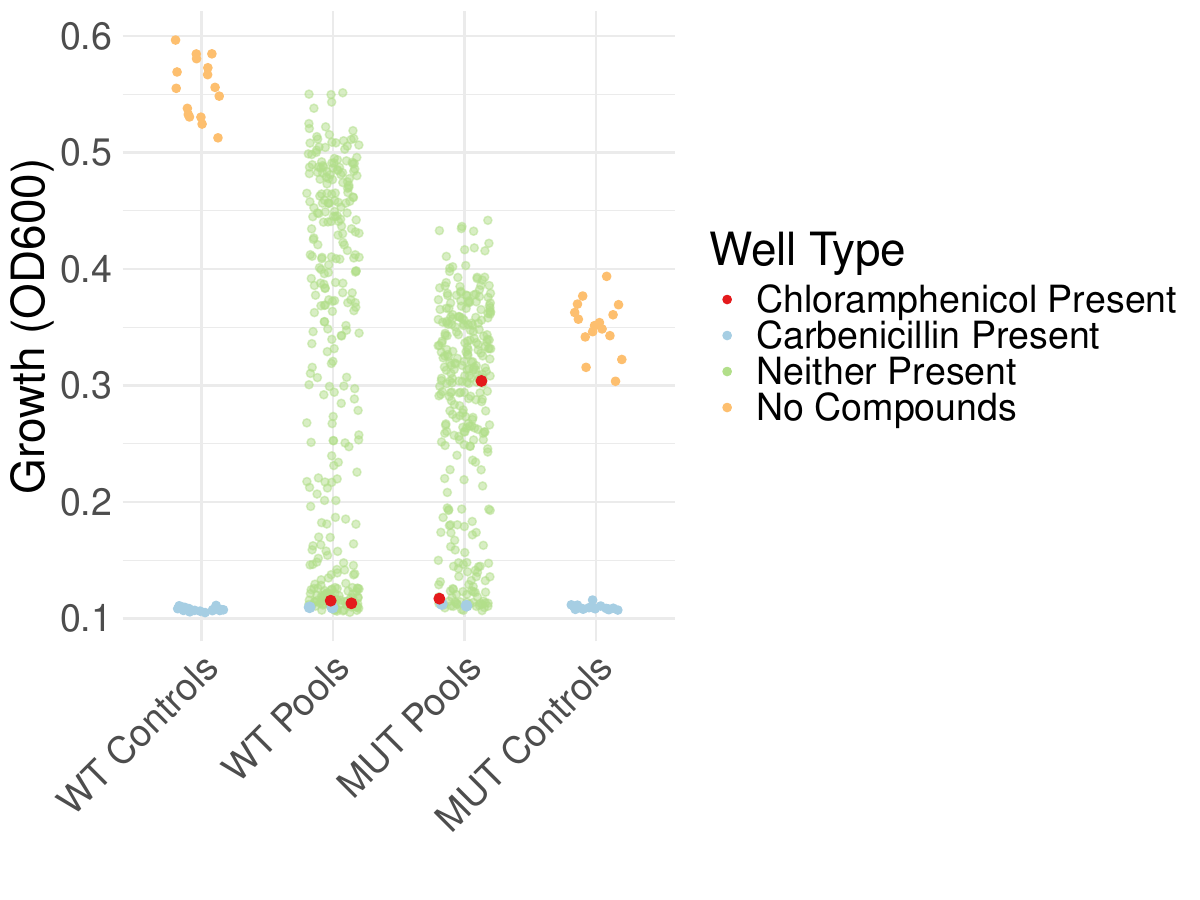}
    \caption{Representation of the amount of cell growth for 32 WT Controls, the 320 WT pooled wells, the 320 MUT pooled wells, and the 32 MUT Controls. The pooled wells studied 1{,}280 compounds. The wells are colored by whether they included the pseudo-hit (Carbenicillin) or the true hit (Chloramphenicol). For the Controls, the Carbenicillin wells included \emph{only} Carbenicillin; for the pooled wells, each well included 8 different compounds.}
    \label{fig:1280}
\end{figure}

Here we provide a plot (Figure \ref{fig:1280}) of the results of the $(n=320, k=1280, c=8)$ proof-of-concept screen. Though there is some visual evidence that the screen identified the pseudo-hit and true hit, we were unable to verify this using the Gauss-Lasso modeling.

\subsection*{Screen Results}

\begin{sidewaysfigure}
  \centering
  \includegraphics[page=1,width=0.48\textheight]{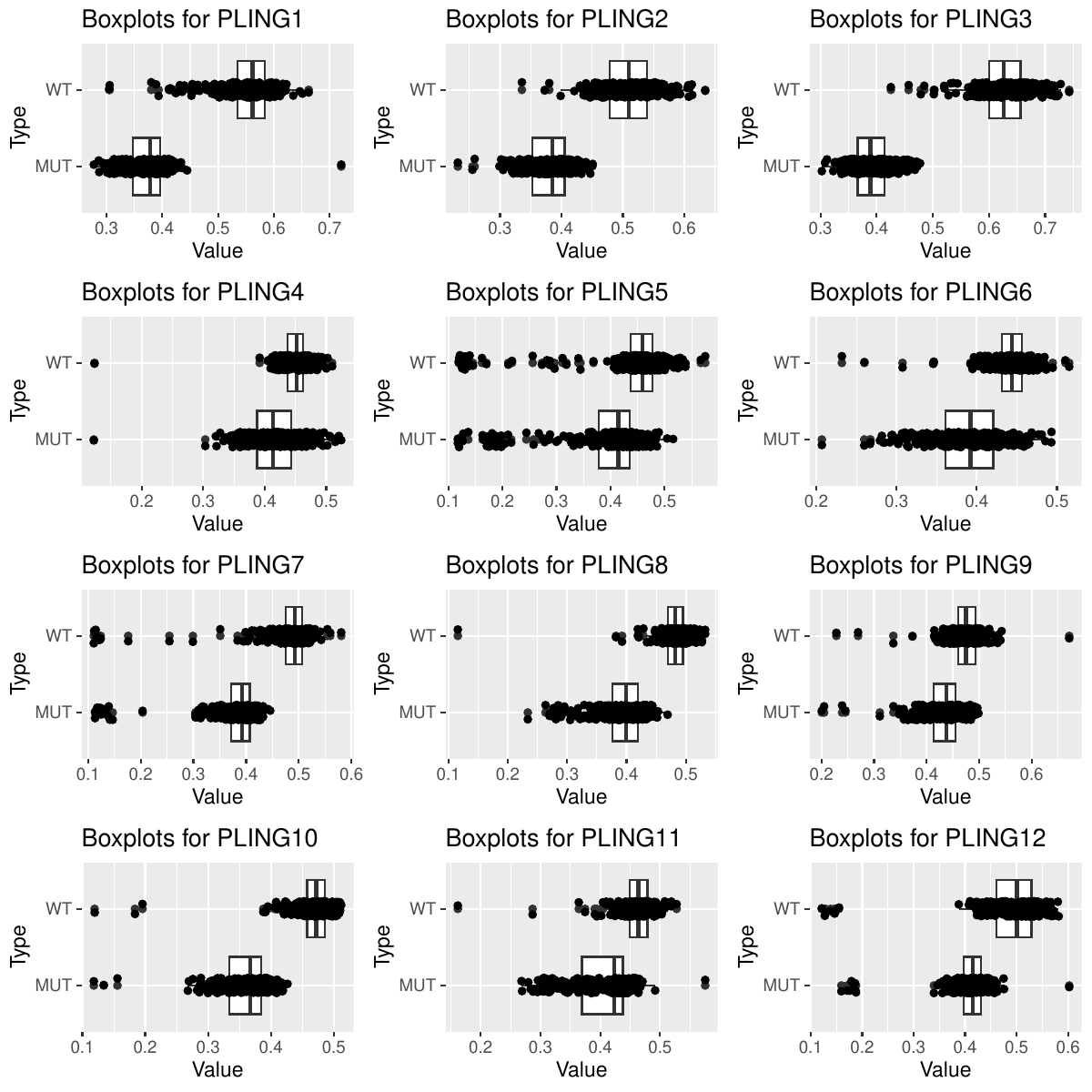}
  \hfill
  \includegraphics[page=2,width=0.48\textheight]{boxplots_PLINGi.pdf}
  \caption{Raw data from the screen described in Section 5.2 of the document.}
    \label{fig:boxplots_PLINGS}

\end{sidewaysfigure}

In Figure \ref{fig:boxplots_PLINGS} we provide boxplots from the small screen described in Section 5.2 in the paper.

\end{document}